\documentclass[sigconf,screen]{acmart}
\usepackage{graphicx}
\usepackage{textcomp}
\usepackage{fancybox}
\usepackage{hyperref}
\usepackage{cleveref}
\usepackage{multirow}
\usepackage{multicol}
\usepackage{booktabs}
\usepackage{pifont}
\usepackage{bbding}
\usepackage{fontawesome}
\usepackage{xinttools}
\usepackage{tabularx}
\usepackage{arydshln}
\usepackage{makecell}
\usepackage{cellspace}
\usepackage{tablefootnote}

\definecolor{log_example_blue}{RGB}{68,114,196}
\definecolor{taint_example_yellow}{RGB}{254,192,0}

\newcommand{\answerbox}[1]{%
    \par\noindent%
    \colorbox{gray!15}{%
        \parbox{\dimexpr\linewidth-2\fboxsep}{%
            \normalsize #1%
        }%
    }%
    \par%
}

\newcommand{\casebox}[1]{%
    \smallskip 
    \noindent 
    \setlength{\fboxrule}{0.1pt} 
    \setlength{\fboxsep}{2pt} 
    \fbox{\colorbox{white}{
        \parbox{\dimexpr\linewidth-2\fboxsep-2\fboxrule}{
            \small\ttfamily #1 
        }%
    }}%
    \smallskip 
}

\AtBeginDocument{%
  }

\copyrightyear{2026}
\acmYear{2026}
\setcopyright{rightsretained}
\acmConference[ICSE '26]{2026 IEEE/ACM 48th International Conference on Software Engineering}{April 12--18, 2026}{Rio de Janeiro, Brazil}
\acmBooktitle{2026 IEEE/ACM 48th International Conference on Software Engineering (ICSE '26), April 12--18, 2026, Rio de Janeiro, Brazil}
\acmPrice{}
\acmDOI{10.1145/3744916.3764570}
\acmISBN{979-8-4007-2025-3/26/04}

\begin{document}

\title{ConfLogger: Enhance Systems' Configuration Diagnosability through Configuration Logging}

\author{Shiwen Shan}
\orcid{0009-0000-1317-8957}
\affiliation{%
  \institution{School of Software Engineering, Zhuhai Key Laboratory of Trusted Large Language Models, Sun Yat-sen University} 
  \city{Zhuhai City}
  \country{China}
}
\email{shanshw@mail2.sysu.edu.cn}

\author{Yintong Huo}
\orcid{0009-0006-8798-5667}
\affiliation{%
  \institution{Singapore Management University}
  \country{Singapore}
}
\email{ythuo@smu.edu.sg}

\author{Yuxin Su}
\orcid{0000-0002-3338-8561}
\affiliation{%
    \institution{School of Software Engineering, Zhuhai Key Laboratory of Trusted Large Language Models, Sun Yat-sen University}
  \city{Zhuhai City}
  \country{China}
}
\email{suyx35@mail.sysu.edu.cn}

\author{Zhining Wang}
\orcid{0009-0005-3214-3350}
\affiliation{%
    \institution{School of Software Engineering, Zhuhai Key Laboratory of Trusted Large Language Models, Sun Yat-sen University}
  \city{Zhuhai City}
  \country{China}
}
\email{wangzhn23@mail2.sysu.edu.cn}

\author{Dan Li}
\authornote{corresponding author}
\orcid{0000-0002-3787-1673}
\affiliation{%
    \institution{School of Software Engineering, Zhuhai Key Laboratory of Trusted Large Language Models, Sun Yat-sen University}
  \city{Zhuhai City}
  \country{China}
}
\email{lidan263@mail.sysu.edu.cn}

\author{Zibin Zheng}
\orcid{0000-0002-7878-4330}
\affiliation{%
    \institution{School of Software Engineering, Zhuhai Key Laboratory of Trusted Large Language Models, Sun Yat-sen University}
  \city{Zhuhai City}
  \country{China}
}
\email{zhzibin@mail.sysu.edu.cn}









\begin{abstract} 
Modern configurable systems, including distributed systems and recently popular decentralized systems in Web 3.0, offer customization via intricate configuration spaces, yet such flexibility introduces pervasive configuration-related issues such as misconfigurations and latent software bugs. 
Existing diagnosability supports focus on post-failure analysis of software behavior to identify configuration issues, but none of these approaches look into \textit{whether the software clues sufficient failure information for diagnosis}.
%
To fill in the blank, we propose the idea of configuration logging to enhance existing logging practices at the source code level.
We develop \texttt{ConfLogger}, the first tool that unifies configuration-aware static taint analysis with LLM-based log generation to enhance software configuration diagnosability. Specifically, our method 1) identifies configuration-sensitive code segments by tracing configuration-related data flow in the whole project, and 2) generates diagnostic logging statements by analyzing configuration code contexts. 

Evaluation results on eight popular software systems demonstrate the effectiveness of \texttt{ConfLogger} to enhance configuration diagnosability.
Specifically, \texttt{ConfLogger}-enhanced logs successfully aid a log-based misconfiguration diagnosis tool to achieve
100\% accuracy on error localization in 30 silent misconfiguration scenarios, with 80\% directly resolvable through explicit configuration information exposed.
In addition, \texttt{ConfLogger}
achieves 74\% coverage of existing logging points, outperforming baseline LLM-based loggers by 12\% and 30\%.
It also gains 8.6\% higher in precision, 79.3\% higher in recall, and 26.2\% higher in F1 compared to the state-of-the-art baseline
in terms of variable logging
while also augmenting diagnostic value. 
A controlled user study on 22 cases further validated its utility, speeding up diagnostic time by 1.25× and improving troubleshooting accuracy by 251.4\%.
    
\end{abstract}





\maketitle
\section{Introduction} 
Modern software systems, including distributed systems and decentralized systems~(e.g., decentralized applications built upon Web 3.0 infrastructure~\cite{vevera2025cybersecurity}), offer configuration options to accommodate the diverse needs of users~\cite{shan2024face,xu2015hey,zhang2014configuration}, where the size of configuration spaces has been growing as systems evolve~\cite{zhang2014configuration,dong2016orplocator}. 
The intricate combinations of configurations, along with insufficient documentation, hinder users from setting up the system and lead to configuration-related issues during implementation, which are known as misconfiguration or configuration bugs~\cite{xu2015hey,keller2008conferr,xu2013not,yin2011empirical,wang2023understanding}. 
Configuration issues can have a wide range of consequences, such as subtle disruptions to the system's functionality~\cite{shan2024face,xu2015hey,li2017conftest}.
In more severe cases, these errors can lead to catastrophic system failures~\cite{xu2013not} and financial losses for big companies~\cite{tang2015holistic}. 
To diagnose misconfiguration, existing studies~\cite{wang2018misconfdoctor,shan2024face, zhou2021confinlog} show superiority in utilizing software operation logs over source code due to the availability of run-time information~\cite{xu2013not,chen2020understanding}.
Taking the following run-time log as an example, its configuration parameter identifier~(i.e., name) \texttt{mapred.local.dir} directly indicates the misconfiguration trigger. \\
\casebox{No valid local directories in property: mapred.local.dir}
Nevertheless, these log-based solutions rely on the system’s \textit{configuration diagnosability}~\cite{yuan2012improving,zhang2015proactive}, i.e., whether the subject system could properly log configuration-related events. 
Furthermore, existing configuration testing tools~\cite{li2024ecfuzz,li2017conftest} are also dependent on the output logs for validity verification.
When different tools are developed to inspect logs, they often overlook a crucial question: 
\textbf{Do existing systems meet the diagnostic requirements for configuration management?}
Configuration diagnosability serves as a basis for upcoming reliable configuration management tools. 

\begin{table*}[t]
\footnotesize
    \centering
    \renewcommand{\arraystretch}{1.5}
        \caption{Motivating examples of deficiencies in configuration diagnosability.}
    \newcolumntype{C}[1]{>{\centering\arraybackslash}p{#1}} 

    \begin{tabularx}{\textwidth}{|c|C{2cm}|c|C{1cm}|>{\raggedright\arraybackslash}X|} 
    \hline
        \textbf{Report ID} & \textbf{Impact} & \textbf{Error-Triggered Conf Param} & \textbf{\# Conf Params} & \textbf{Explanation} \\
    \hline
        HCommon-14604 & Silent failure & \texttt{dfs.replication} & 398 & 
        Silent failure causes the system to ignore \texttt{dfs.replication} in \texttt{hdfs-site.xml}, forcing users to manually check replication counts due to no warnings/errors. \\
    \hline
        HDFS-2820 & Insufficient diagnostic messages & 
        \makecell{\texttt{dfs.namenode.shared.edits.dir} \\ \texttt{dfs.namenode.rpc-address}} & 93 & 
        Insufficient diagnostic messages omit the misconfigured namenode address when configuring the shared edits directory, causing user confusion and prolonged debugging. \\
    \hline
    \end{tabularx}
    \label{tab:motivating_examples}
\end{table*}

Based on existing studies~\cite{sun2020testing,xu2015hey,zhang2021static}, we identify two configuration deficiencies in contemporary systems, categorized by\textit{silent failures} and \textit{insufficient diagnostic messages}. 
Table~\ref{tab:motivating_examples} presents two real-world user reports of such configuration limitations and their impacts.
Silent failures indicate scenarios when parameter value conflict occurred with undocumented fallback (i.e., no clear anomalous output)~\cite{zhang2021static}. In the first case\footnote{\url{https://issues.apache.org/jira/browse/HADOOP-14604}}, Hadoop fails to produce warning logs on replication checking failure, which might obviate users to validate system behavior and cause another failure later.
Insufficient diagnostic messages denote scenarios when system outputs do not explicitly locate misconfigured parameters by name or value~\cite{shan2024face,zhang2015proactive,xu2013not}, which hinders users from rectifying misconfiguration.
For example, the second report\footnote{\url{https://issues.apache.org/jira/browse/HDFS-2820}} displays the insufficient system feedback on problematic configuration parameters, preventing users from diagnosing namenode errors. 

To promote the diagnosability of software systems, we propose a \textit{configuration logging} strategy that improves configuration-related logging practices within the existing codebase. The insight is that developers should proactively write down logging statements outlining potential misconfiguration causes based on the surrounding code context. In this way, users can leverage run-time logs to troubleshoot configuration issues without having to inspect the source code~\cite{shan2024face}. 


Enhancing configuration diagnosability through configuration logging is challenging in: 
(1) detecting configuration-sensitive code segments that could have impact on software behavior,
and (2) generating readable and meaningful logs for diagnostics. 
The first challenge stems from the diverse utilization of configuration parameters in the whole software project, spanning across declaration, propagation, and usage phases~\cite{chen2020understanding,wang2023conftainter}. 
Since the declaration and usage phases are decoupled, 
we need to track parameter propagation to identify scattered usage phases, thereby locating configuration-sensitive code segments that encapsulate critical constraints and validation logic for logging.
The second challenge involves optimizing logging contents in describing misconfiguration symptoms and underlying reasons. 
Diagnostic logs should include essential configuration information~\cite{zhang2015proactive,shan2024face}, including identifiers, 
runtime values, 
and their causal relationships with system behavior. 
This requires mechanisms to dynamically map configuration states to operational impacts and encode these relationships through structured log formats. 
%

To boost system configuration diagnosability via logging support, we present \texttt{ConfLogger}, including a configuration-sensitive code identification component and a logging statement generation component. The two components resolve the abovementioned two challenges, respectively.
Specifically, in the first component, \texttt{ConfLogger} starts with labeling the configuration classes in the codebase by mapping from configuration documents. 
Then, it tracks data flow and control flow information of the configuration parameters to localize the logical entry points of configuration usage code by taint analysis.
These entry points, usually being configuration parameters checking statements~(e.g., branches), 
provide context on how particular configuration parameters are used and impact system behavior in source code.
Finally, \texttt{ConfLogger} extracts code segments associated with these tracked entry points.
%
%
In the second component, we leverage Chain of Thought (CoT)~\cite{wei2022chain}-enhanced LLM to automatically generate logging statement contents for the configuration-sensitive code segments. 
Firstly, LLM analyzes existing code to determine the necessity to inject new logging statements. 
Then, it determines the optimal log points based on block features and handcrafted logging instructions,
and generates complete logging statement accordingly.
In a nutshell, our approach gains the advantages of project-level configuration tracing via code analysis, and context-aware log generation via language models, enabling precise diagnosis without manual instrumentation.


To evaluate the effectiveness of \texttt{ConfLogger} in improving system diagnosability, we investigate whether the injected logs enhance the capability of the state-of-the-art log-based misconfiguration diagnosis tool across eight representative systems.
Experimental results show a 100\% diagnosis accuracy when using \texttt{ConfLogger}-enhanced logs upon 30 silent misconfiguration cases, compared to 0\% original accuracy due to silent failures. 
80\% of cases are directly localized via configuration parameter names/values exposed in logs (e.g., "Please set ‘mapreduce.framework.name’ as ‘yarn’").
Moreover, compared with other LLM-based logging baselines,
\texttt{ConfLogger} outperforms state-of-the-art solutions by +8\% (74\% vs. 66\%) and +17\% (74\% vs. 57\%) in the coverage of existing log points
and +8.6\%/+51.5\% higher in precision, +79.3\%/+138.3\% higher in recall and +26.2\%/+42.3\% higher in F1 score in variable logging
with enriched semantics of logging statements.
Regarding efficiency, \texttt{ConfLogger} achieves automated identification of configuration entry points, outperforming manual approaches that require 154.696 seconds of human effort to analyze source code statements and extract validation rules, with a 39.36× speedup and a 66.5\% reduced invalid rate in our experiments.
Last but not least, user studies of \texttt{ConfLogger} demonstrate its practicality by speeding up 1.25× diagnosis time and increasing diagnosis accuracy by 251.4\% for configuration failures. 

In summary, our main contributions are listed as follows:

$\blacklozenge$ To the best of our knowledge, we are the first to introduce configuration logging as a proactive strategy for enhancing configuration diagnosability.

$\blacklozenge$ We design and implement \texttt{ConfLogger}\footnote{\url{https://github.com/shanshw/ConfLogger}}, to enhance existing logging practice by two components, i.e., configuration-sensitive code identification and configuration logging statement generation.

$\blacklozenge$ We demonstrate the logging quality of \texttt{ConfLogger} on eight software systems, and further show its practicality through user study. 

\section{Problem Definition}
In this paper, we formulate the \textit{configuration logging} problem as follows.
Given a target system's code $c_{t}$,
a logging tool should locate its configuration-sensitive code $c_c$, and then outputs its enhanced version $c_e$ with $q$ newly injected configuration-related logging statements $s_{log}$.
The relation between $c_c$ and $c_e$ is: $c_{e} = c_{c} \cup \{ s_{log1}, ..., s_{logq}\}$.

In particular, we decouple the task into two sub-tasks.
The first sub-task, is to locate the configuration-sensitive code segments $c_c$. 
With target system's code $c_{t}$ containing $n$ statements $s$, $c_{t} = \{ s_1, ..., s_n \}$, 
we locate configuration-related code segments $c_{c}$ with $m$ configuration-related statements $s_{c}$, 
$c_{c} \in c_{t}$, $ \{s_{c1}, ..., s_{cm} \} \in c_c$.
Taking $c_c$ as input,
the second sub-task targets generating configuration-informative logging statements
$s_{log}$ and inserting them into $c_c$, and finally outputs the enhanced code $c_{e}$.
To guarantee $s_{log}$ is configuration-informative, $s_{log}$ should contain following information of configuration parameters:
(1) Configuration parameter identifiers~(i.e., names) or configuration parameter values, referring as configuration variables $var_{c}$,
(2) Configuration constraints~(e.g., configuration dependencies, numeric ranges, etc.)~\cite{chen2020understanding,xu2013not} $text_{cc}$, and
(3) Configuration setting guide $text_{cg}$ for potential misconfiguration resolution.
Therefore, we get:
$\{var_{c1}, ... var_{cx} \} \cup \{text_{cc1}, ... text_{ccy} \} \cup \{ text_{cg1}, ..., text_{cgz} \} \in s_{log}$.

\section{Related Work}

\textbf{Configuration Practices.}
Misconfiguration occur when configuration setting violate the constraints~(e.g., invalid value ranges, unsatisfied parameter dependencies,etc.)~\cite{zhang2021static,zhang2014configuration,liao2018you,yin2011empirical}. 
To extract configuration constraints, 
SPEX~\cite{xu2013not} applies static analysis with predefined rules to track configuration parameter data flow.
CDep~\cite{chen2020understanding} empirically identifies configuration dependencies through static analysis,
while ConfInLog~\cite{zhou2021confinlog} infers constraints directly from log messages.
For misconfiguration diagnosis, 
ConfDiagnoser~\cite{zhang2013automated} combines static/dynamic analysis with statistical components.  
ConfigX~\cite{zhang2021static} localizes silent misconfigurations via static program analysis and deep relation inference.
Ciri~\cite{lian2024large} leverages few-shot LLM technology to locate misconfigurations using historical data,
while LogConfigLocalizer~\cite{shan2024face} employs rule-based log analysis augmented with LLMs.
MisConfDoctor~\cite{wang2018misconfdoctor} proactively injects misconfigurations to derive diagnostic signatures.
Prior work assesses misconfiguration diagnosability ~\cite{yin2011empirical,xu2015hey,xu2013not} through configuration error injection testing (CEIT)~\cite{li2021challenges, li2018confvd,li2017conftest,keller2008conferr,li2024ecfuzz}, which injects errors into systems and analyzes their responses. CEIT tools include CeitInspector (a systematic evaluation framework)~\cite{li2021challenges},
ConfErr~\cite{keller2008conferr}, ConfVD~\cite{li2018confvd} and ConfTest~\cite{li2017conftest} 
which define injection rules and verify validity via system outputs (e.g., logs).
ConfDiagDetector~\cite{zhang2015proactive} enhances diagnosability assessment by combining injection with NLP-based log analysis. In contrast, PCHECK~\cite{xu2016early} proactively improves system reliability by injecting preemptive parameter checks.

Existing tools address misconfigurations effectively but rely on high system diagnosability, while CEIT tools lack systematic strategies to enhance diagnosability, failing to meet user and tooling requirements
~\cite{xu2013not,xu2015hey,zhang2015proactive}.
In contrast, \texttt{ConfLogger} enhances configuration logging by explicitly exposing parameter identifiers and other critical information, thereby fundamentally improving diagnosability.

\textbf{Logging Practices.}
Logging research spans upstream works (logging statement generation~\cite{yuan2012improving,li2024go,xu2024unilog,ding2012ErrLog,jia2018smartlog},
logging bug detection~\cite{zhong2024automated,chen2019extracting,chen2017characterizing}), midstream works (log parsing~\cite{he2017drain,huo2023semparser,jiang2024lilac})
, and downstream works (anomaly detection~\cite{huo2023evlog,du2017deeplog}, 
root cause analysis~\cite{shan2024face,wang2018misconfdoctor}).
Logging statement practice can be further divided into two categories based on their objectives: 
1) enhancing the quality of existing logging code and 2) providing automatic suggestions on logging-free code.
The first category focuses on improving existing logging's capability in error-handling scenarios, thereby promoting systems' reliability~\cite{yuan2012improving, ding2012ErrLog,jia2018smartlog}.
The second class automatically generates logging statements learning from developers' logging history in general-purpose scenarios~\cite{li2024go,li2020towards,zhu2015learning,liu2022tell,he2021survey}. 
These studies contain decisions-making in logging points~\cite{li2020shall,yao2018log4perf,zhao2017log20} (e.g., log points at the line level when given target methods), logging levels~\cite{li2017log,li2021deeplv,liu2022tell} (e.g., verbosity level prediction), and constructing complete logging statements~\cite{li2024go,xu2024unilog,mastropaolo2022using}.
While these studies provide automated logging suggestions, the learning-based nature blocks them from overcoming current drawbacks on logging quality~\cite{li2020qualitative,zhu2015learning,li2021deeplv}.

\texttt{ConfLogger} belongs to the first category but uniquely locates configuration-sensitive code and enhances log instrumentation with configuration-specific diagnostic data. 
Its diagnosability-centric approach and logging quality improvement further distinguish it from second-category works.
\section{ConfLogger}
\subsection{Overview} 
\begin{figure}[tbp]
\small
    \centering
\includegraphics[width=0.8\linewidth]{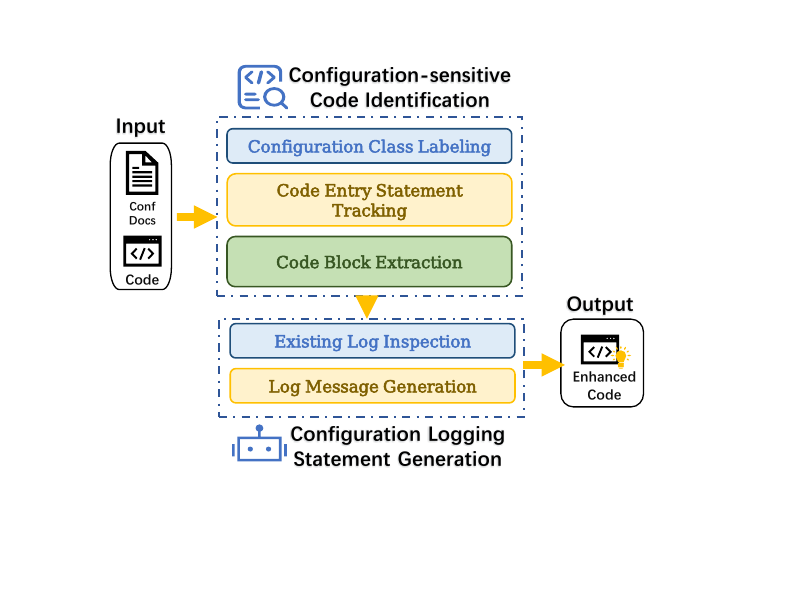} 
    \caption{Overview of \texttt{ConfLogger}.}
    \label{fig:overview}
\end{figure}
\begin{figure*}[htbp]
    \centering
    \includegraphics[width=0.9\linewidth]{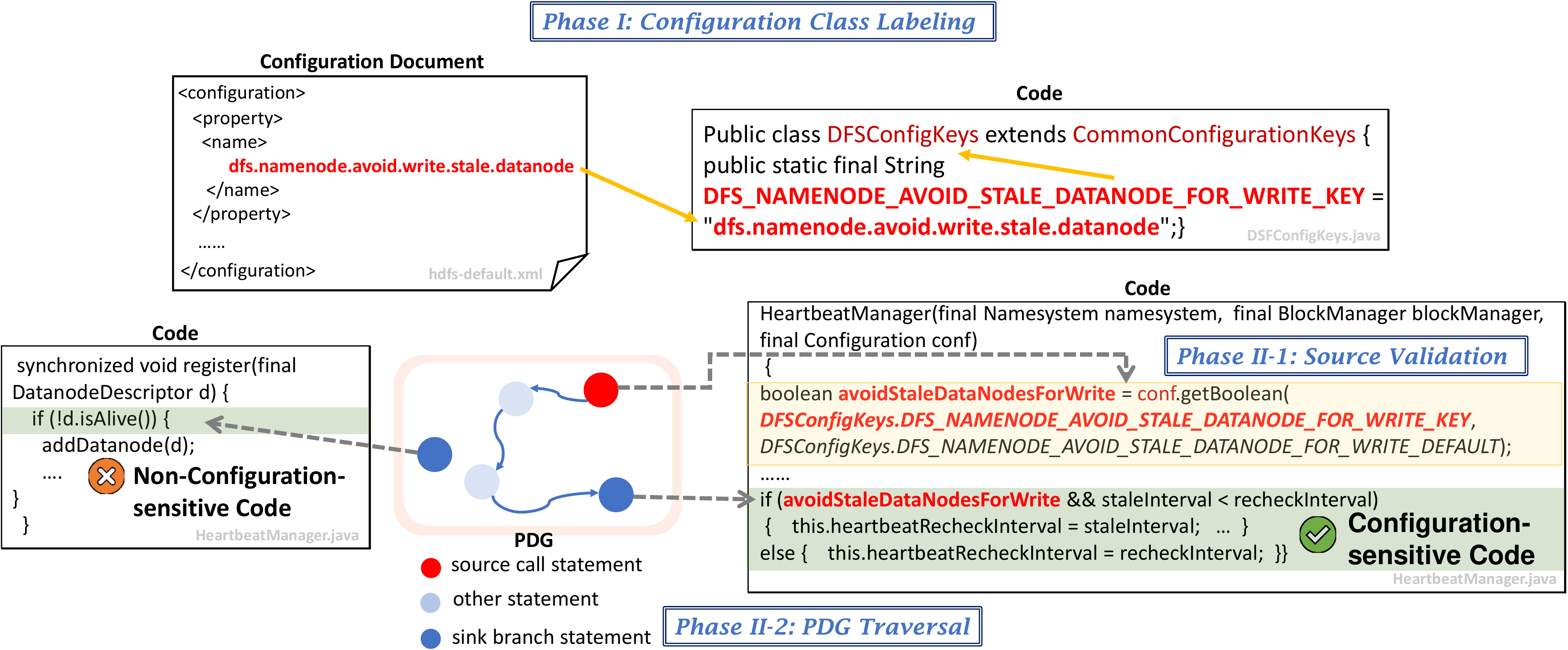}
    \caption{An example of configuration-sensitive code identification. The red text indicates objects colored by the configuration parameter, thus being configuration-related.}
    \label{fig:taint_example}
\end{figure*}
Fig.~\ref{fig:overview} depicts the design of \texttt{ConfLogger}. Given the system's code and configuration documentation as input, 
\texttt{ConfLogger} outputs enhanced code with injected logging statements.
Specifically, our model is composed of two components, i.e., configuration-sensitive code identification and configuration logging statement generation.
The first identifies configuration-sensitive code blocks as logging target code, and the second injects informative logging statements on the identified code.
In the first component, \texttt{ConfLogger} starts with labeling all configuration engine classes in the source code based on the official documentation.
Then, it tracks the statements with configuration parameters along the whole software system based on the labeled classes.
Based on these statements, \texttt{ConfLogger} further extracts the associated code contexts, denoted as configuration-sensitive code segments.
The second component, empowered by LLMs, takes these configuration-sensitive code segments for logging. In particular, \texttt{ConfLogger} firstly determines the necessity of inserting new logging statements. Then, we ask LLMs to generate complete logging statements based on instructions and a set of configuration-tailored logging rules. 
The output of \texttt{ConfLogger} is the enhanced code equipped with configuration logging.

\subsection{Configuration-sensitive Code Identification}
Considering the large size of the project-level codebase, logging every variable at runtime is impractical and could significantly impact software performance. Therefore, 
this component begins with identifying scattered code blocks that contain configuration usages.

\subsubsection{Code Unit for Configuration Logging}
\textit{Configuration checking} and \textit{handling} are widely used in existing configuration practices~\cite{chen2020understanding,zhang2021static,xu2016early}. 
In particular, the checking part validates configuration constraints (e.g., ensuring parameter values fall within predefined ranges, 
verifying relationships between configuration parameters, 
and detecting any missing required parameters~\cite{xu2016early,chen2020understanding,liao2018you}), and the handling part processes these validation results and mitigates them accordingly.
Following these practices, we choose code segments that involve the configuration checking-handling mechanism as \textit{configuration-sensitive code} unit. Logging over these configuration-sensitive units further offers two advantages:
1) capture system behavior: varying configuration settings can lead to different system behaviors within these code blocks, and 
2) provide configuration contexts: the checking-handling process provides essential context for capturing configuration intentions.


An example of configuration-sensitive code is illustrated in the lower corners of Fig.~\ref{fig:taint_example}, where
the code block marked with a green background indicates the checking logic and the handling logic.
In this case,
the checking part checks if the configuration variable \texttt{avoidStaleDataNodeForWrite} is set to ``true'' and ensures the other two configuration variables (i.e., \texttt{staleInterval} and \texttt{recheckInterval}) satisfy their dependency relationships.
The handling part then assigns \texttt{this.heartbeatRecheckInterval} the value of either \texttt{staleInterval} or \texttt{recheckInterval} based on the checking results. 

This example illustrates that such code contains rich contextual configuration information, 
including the values of configuration variables and their interdependencies.
Such code context can hint at configuration errors, thereby providing feasible insights for troubleshooting.
In contrast, the left panel shows another example that will not be seen as configuration-sensitive code.
Here, the green-highlighted branch implements the checking logic, but the condition variable \texttt{d.isAlive()} is configuration-independent, thus failing to clue configuration information.

\subsubsection{Configuration Class Labeling}
Identifying configuration-sensi-tive code is challenging for two reasons:
first, a software system may have large codebases with multiple files, where configurations might spread across various locations. These files may exceed LLMs' context length to analyze code and locate certain blocks.
Second, the presence of multiple intricate components across various classes, along with complex inheritance hierarchies and composition relationships, hinders comprehensive coverage of configuration engine classes. As a result, manual identification on configuration engines fails to meet practical requirements.

To overcome these challenges, 
we begin by automatically labeling configuration engine classes from configuration documents.
The intuition here is to color every object and class with configuration parameters.
In particular, \texttt{ConfLogger} extracts parameter keys from documentation to map their identifiers in source code.
These mapped identifiers are then colored by the extracted parameters.
\texttt{ConfLogger} identifies seed configuration classes containing these colored identifier variables as filed members.
Then, it expands these classes through inheritance hierarchies and composition relationships (e.g., inner and anonymous classes) to label the final configuration engine classes.
To ensure comprehensiveness, we incorporate both Java built-in structures and user-specified configuration engines.

The upper side of Fig.~\ref{fig:taint_example} displays an example, where
\texttt{ConfLogger} parses the configuration document and extracts the configuration parameter name, \texttt{dfs.namenode.avoid.write.stale.datanode} as a unified identifier. 
Then, it maps the configuration variable\footnote{\texttt{DFS\_NAMENODE\_AVOID\_STALE\_DATANODE\_FOR\_WRITE\_KEY}}
to the unified qualifier indicated by the yellow arrow. 
Since the configuration variable is defined as filed members in the java class \texttt{DFSConfigKeys},
we label the class as a seed configuration engine.
Besides,
\texttt{ConfLogger} includes \texttt{CommonCOnfigurationKeys} as an expanded configuration engine due to the hierarchy relations.

\subsubsection{Code Entry Statement Tracking}
To locate configuration-sensi-tive code, we trace data and control flows from configuration parameter definitions to their usage. The logical entry statement in each usage instance would be considered the starting point of configuration-sensitive code.
To this end, we conduct inter-procedural analysis on the Program Dependence Graph (PDG) to track all method call edges and path-sensitive data flow based on the idea of taint analysis.
PDG is a directed graph representation that
encodes data/control dependencies (edges) between program statements (nodes)~\cite{ferrante1987program}. 

To track along PDG, we specify all call statements that are associated with the getter methods in the labeled configuration engines, referred to as candidate source statements for tracing. 
These getter methods conventionally manage hierarchical key-value mappings. Existing techniques~\cite{chen2020understanding, xu2015hey} rely on manual annotation of configuration engines to derive source statements, resulting in limited coverage and requiring expertise.
To address this, we propose an automated approach to identify valid source statements out of the candidates by a type-specific validation mechanism to getter methods.
Specifically, we tailor rules for different types of configuration engines~(Table~\ref{tab:getter-method-rules}):
(1) We exclude call statements to the getter methods of \texttt{Key-Holder}~(storing parameter identifiers only) as it doesn't store the parameter values that can taint variables;
(2) We enable untyped parameter (no constraints) on \texttt{Both-Holder} since it combines key-based and dictionary-based mechanisms, allowing parameter access via direct keys and built-in getter methods for specific parameters~(e.g., \texttt{getResilient()}).
Doing so offers balanced flexibility.
(3) We propose constrained parameter types (\texttt{Key-Holder} or \texttt{Both-Holder} only) on the getters of \texttt{Dict-Holder}~(managing configuration dictionaries), reducing false positives through type-restricted value access.
This strategy ensures valid source statement identification through qualified getter calls while preserving semantic alignment with configuration management patterns. 

\begin{table}[t]
\small
  \caption{Parameter type requirements on different configuration engine types.}
  \label{tab:getter-method-rules}
    \begin{tabular}{|c|c|} 
    \hline 
    {\textbf{Parameter Type}} & {\textbf{Configuration Engine Type}} \\ 
    \hline 
    \texttt{{Unified Identifiers}} & \texttt{/}    \\ 
    \cline{1-2}
    \texttt{/} & \texttt{Both-Holder}  \\
    \cline{1-2}
    \multirow{2}{*}{\texttt{Key-Holder} \& \texttt{Both-Holder}} & \multirow{1}{*}{\texttt{Built-in}}  \\ 
    {} & \texttt{Dict-Holder} \\ 
    \hline
  \end{tabular} 
\end{table}
\begin{figure*}[t] 
    \centering
    \includegraphics[width=\linewidth]{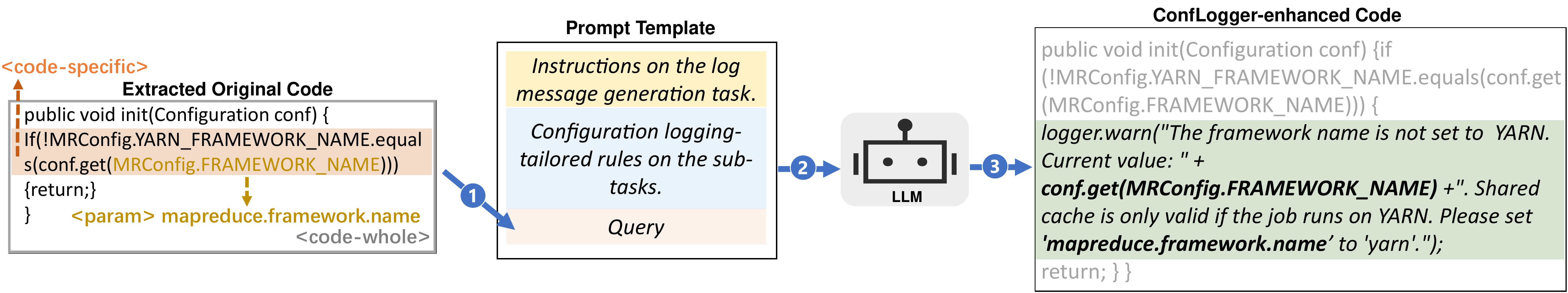}
    \caption{An example case of logging statement generation.}
\label{fig:logging_statement_generation_examle}
\end{figure*}
Sink statements are the destinations of the tracking phase, namely the usage points of configuration parameters.
Therefore, we leverage configuration taint analysis to achieve tracking from the identified source statements to these sink statements.
Configuration taint analysis is a program analysis approach that tracks the propagation of configuration parameters through execution paths~\cite{fu2024missconf,wang2023conftainter}, thereby helping tracking configuration usages~(i.e., tainted sink statements in technical concepts) in a complicated codebase along the propagation of configuration-related variables.
In particular, 
\texttt{ConfLogger}
establishes a PDG for the codebase.
The PDG is constructed based on Static Single Assignment Intermediate Representation (SSA IR), enabling precise tracking of both data and control dependencies across procedures.
Then it employs the Breadth-First Search Algorithm with limited path lengths to traverse the PDG, tracking tainted sink statements.
The path length constraint ensures LLMs' stable capability.

Phase II in Fig.~\ref{fig:taint_example} presents the detailed workflow.
In Phase II-1,
\texttt{ConfLogger} firstly validates the call statement pointed out by the source node~(in red circle) in the established PDG since it's a call statement to the public getter method~\texttt{getBoolean} of \texttt{Both-Holder} configuration engine, satisfying the validation rules.
And then, \texttt{ConfLogger}
labels the variable \texttt{avoidStaleDataNodesForWrite} as taint configuration variable because it carries the return value of the identified source statement.
In Phase II-2,
\texttt{ConfLogger} tracks the taint's data and control flow on the PDG and finds a path reaching a sink branch statement within three hops pointed by the PDG's sink node~(in dark blue).
\texttt{ConfLogger} thus detects the tainted sink branch statement, which achieves the checking logic of configuration-sensitive code.
These located statements implementing configuration-checking logic will later aid in extracting configuration-sensitive code segments.

\subsubsection{Code Block Extraction}
The located taint sinks are the checking part of configuration-sensitive code.
To include their corresponding handling part, 
we start with grouping the tainted statements by methods.
We achieve this by classifying them based on their method signatures. 
Methods containing these statements are identified as configuration methods and are provided as context to the LLM during subsequent logging statement generation phase.
Then, we leverage the WALA framework~\cite{wala} to obtain the line numbers of these SSA IR statements in the source code, which allows us to determine the starting line numbers of the checking part and extract the handling code accordingly.

\subsection{Configuration logging statement Generation}
As LLMs have demonstrated strong capabilities in code understanding and generation~\cite{li2024go}, we prompt LLMs with Chain-of-Thought (COT) technologies to generate configuration-related logging statements.
Given the configuration-sensitive code segments and the tracked parameters as input, the LLM is required to determine whether to enhance current logging statements or not, followed by identifying optimal log positions and generating the corresponding logging statements.
Ultimately, the LLM outputs the enhanced code segments. 
As illustrated in Fig.~\ref{fig:logging_statement_generation_examle},
in Step {\textbf{\textcolor{log_example_blue}{\ding{182}}}}, \texttt{ConfLogger} marks three key features in the configuration-sensitive code block: 
\texttt{<code-specified>} (tagging the entry point of the extracted code), \texttt{<code-whole>} (capturing the surrounding code context), and \texttt{<param>} (specifying the configuration parameter \texttt{mapreduce.framework.name}).
In Step {\textbf{\textcolor{log_example_blue}{\ding{183}}}}, we assemble these code features to construct a prompt for LLM-based logging generation via CoT reasoning. 
Finally, Step {\textbf{\textcolor{log_example_blue}{\ding{184}}}} produces enhanced code with a newly-injected logging statement. 
The generated log message includes the configuration variable~\footnote{\texttt{conf.get(MRConfig.FRAMEWORK\_NAME)}} and actionable guidance~\footnote{Please set ‘mapreduce.framework.name’ to ‘yarn’} to enable the \texttt{Shared} \texttt{Cache} module in MapReduce.
\subsubsection{Existing Log Inspection}
To mitigate excessive logging, we systematically evaluate existing logging statements on the extracted code following established practices~\cite{yuan2012improving}. 
We establish rule-based criteria for LLM decision-making: When the given code contains logging statements with sufficient configuration details (e.g., parameter keys or value ranges), the LLM recommends retaining existing logs; otherwise, it suggests removing redundancies while annotating rationale based on runtime behavior. 
The LLM skips inspection for code without existing logging statements.

\subsubsection{Log Message Generation.}
In this step, LLMs are required to determine the optimal position in code blocks and produce contextual logging statements accordingly.
To control logging overhead on selecting essential path constraints, we define rules for different checking scenarios:
(1) Insert logs on paths handling invalid/unset values to expose silent default fallbacks as former works points out~\cite{xu2013not,zhang2021static}, therefore revealing constraint violation corrections;
(2) Instrument service activation/deactivation paths to verify configuration prerequisites as the second motivating example on Table~\ref{tab:motivating_examples} exposes, hence guaranteeing intended behavior, and confirming service-switching outcomes;
(3) Insert logs on configuration processing paths to monitor parameter transformations, and validate value-driven system behavior.
The observation is that this information uncovers the hidden configuration management in code, therefore delivering troubleshooting guidance for potential misconfigurations.
Furthermore, we ask LLMs to generate log contents following SLF4j standards~\cite{slf4j} with three mandatory components: 
(1) severity levels selected from predefined SLF4j options
(2) static messages embedding configuration constraints for diagnostics and
(3) dynamic variables capturing parameter names and runtime values, aligned with configuration troubleshooting practices~\cite{shan2024face,xu2013not,zhang2015proactive}. 
Additionally, the LLM was also instructed to offer configuration guidance to mitigate potential misconfigurations.
\begin{table}[t]
    \small
  \caption{Target systems. {*} The version is 3.0.0-beta-1}
  \label{tab:systems}
    \begin{tabular}{c|c|c|c} 
    \toprule 
    {Systems} & {Version} & {LoC} & {\# Configuration Parameters}  \\ 
    \hline
    \texttt{{Storm}} & 2.6.2 &208,677 & 275 \\
    \texttt{{Hbase}}$^{*}$ & {3.0.0} & 87,438& 232 \\
    \texttt{{Alluxio}} & 3.1.3 & 78,541& 821 \\
    \texttt{{HCommon}} & 3.3.6 & 180,161 & 461 \\
    \texttt{{Mapreduce}} & 3.3.6 & 122,822& 223 \\
    \texttt{{Yarn}} & 3.3.6 & 323,532& 545 \\
    \texttt{{HDFS}} & 3.3.6 & 557,733& 643 \\
    \texttt{{ZooKeeper}} & 3.9.2 & 62,681 & 167 \\
    \bottomrule 
  \end{tabular} \\
\end{table}
\begin{figure*}[htbp]
    \centering
    \includegraphics[width=\linewidth]{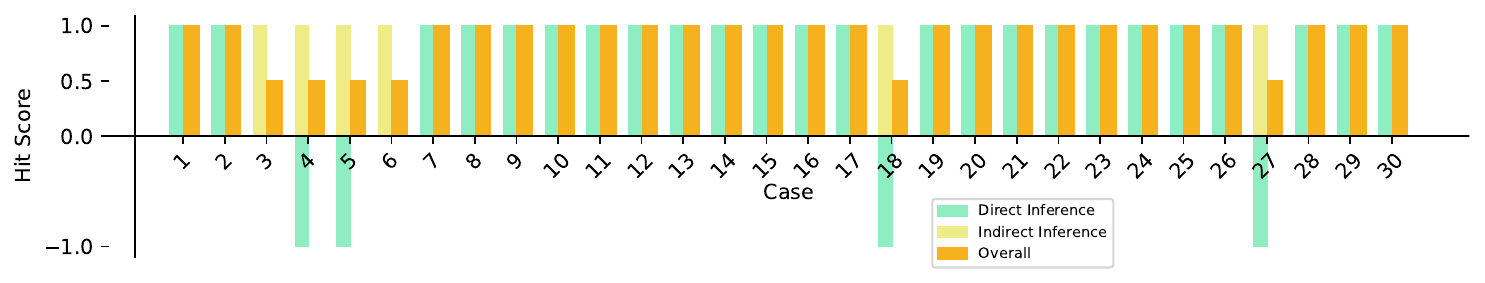} 
    \caption{Overall Hit Scores and Hit Scores of different inference phases.}
    \label{rq1:mis-result}
\end{figure*}
\section{Evaluation}
We conduct experiments for the following research questions~(RQs):

    \textbf{\textit{RQ1}: How effective of \texttt{ConfLogger} on enhancing systems' configuration diagnosability?}

    \textbf{\textit{RQ2}: How does \texttt{ConfLogger} comparing other logging tools on configuration logging?}
              
    \textbf{\textit{RQ3}: How would the source identification strategy affect logging?}
    
    \textbf{\textit{RQ4}: To what extent does ConfLogger help users in misconfiguration diagnosis? (Practical user study)}

\subsection{Experimental Setup}
We implement \texttt{ConfLogger} by a Java-based program analysis module with 4,000 LOC using WALA~\cite{wala} and ASM~\cite{asm}, 
and an LLM interface invoking GPT-4o-2024-08-06 via OpenAI API~\cite{gpt-4o} with temperature set to 0
for reproducibility. 
Following prior studies on configuration practice~\cite{shan2024face,xu2015hey,sun2020testing,xu2019configuration,xu2016early}, we evaluate our method in eight prominent Java systems:
Storm, HBase, Alluxio, Hadoop Common, MapReduce, Yarn, HDFS, and ZooKeeper. Table~\ref{tab:systems} details these systems, with their code statistics calculated using the cloc tool~\cite{cloc}.

We collect JAR/source code files of target systems from the Maven repository~\cite{maven} and use official configuration documentation as input for \texttt{ConfLogger}. 
The average execution time is 355.5 minutes. To balance efficiency and accuracy, we constrain tainted path length to 30, mitigating redundant long-path exploration.
All experiments ran on an x86\_64 server (80-core Intel Xeon Gold 5218R @2.10GHz, 267GiB RAM, 5.6TB disk) through Docker on Ubuntu 22.04, with Python 3.12.4 and Java 11.
\subsection{Datasets}
Benchmark I includes 30 enhanced code cases whose original implementation suffers from silent failures.
These cases' original code fails to log critical configuration-related information, including failures to activate a service due to neglect of certain configuration parameters, out-of-valid-range value settings, thus expose severe potential misconfiguration risks. 
We identify these case from its original implementation and employ \texttt{ConfLogger} to replicate them.
Therefore, they serve as the perfect experiment benchmark to highlight \texttt{ConfLogger}'s capability on enhancing configuration diagnosability, with silent failures eliminated.

Benchmark II includes 70 original configuration-sensitive code cases with 90 log points removed. These log points, identified by \texttt{ConfLogger} as diagnostically sufficient and manually validated as configuration-related, serve as ground truth to evaluate \texttt{ConfLogger}-enhanced logs against two LLM-based baseline loggers in subsequent comparisons.
\subsection{Evaluation Results}
\subsubsection{\textbf{RQ1: How effective of \texttt{ConfLogger} on enhancing systems' configuration diagnosability?}} 
We validate the effectiveness of \texttt{ConfLogger} on Benchmark I using LogConfigLocalizer~\cite{shan2024face} by measuring successful case located by the tool.

\textit{\textbf{Experiment Setting.}}
We collect run-time logs by executing test suites on original and \texttt{ConfLogger}-enhanced code in Benchmark I.
For untested cases, we manually craft test methods.
We employ a log-based misconfiguration diagnosis tool to locate the misconfiguration-trigger parameters. 
LogConfigLocalizer employs rule-based extraction of explicit parameters in logs during its direct inference phase, validated via LLM verification. 
If the direct inference fails, it activates the indirect inference, using LLMs to deduce implicit configuration issues from ambiguous logs.

\textit{\textbf{Metrics.}}
We propose \textit{Hit Score}, awarding 1 (direct success), 0.5 (indirect success after failure) or 0 (total failure) for overall diagnosis, with phase-specific scores of 1/0/-1. 
Higher scores indicates stronger diagnosability of \texttt{ConfLogger} logs.

\textit{\textbf{Results.}}
Fig.~\ref{rq1:mis-result} presents the results. 
All of the 30 cases received an Overall Hit Score btween 0.5 and 1, confirming successful localization in at least one phase.
This yields 100\% localization accuracy, contrasted against original implementations' silent failures. 
Six cases scored 0.5 due to insufficient explicit information (e.g., missing parameter names/values), with four root causes:
(1) Misleading parameter-like tokens (4 cases),  
(2) Logged variable names instead of documented constants (2 cases), 
(3) Ambiguous natural language descriptions (2 cases),  
and (4) Parameter name/documentation inconsistencies (2 cases).  
While these limitations stem from documentation deficiencies or LLM hallucinations~\cite{ji2023survey,rahman2024code}, all achieve Indirect Hit Scores of 1, demonstrating latent diagnosability.  
The majority (24/30) earn direct success (Hit Score 1), indicating enhanced logs' effectiveness in explicit parameter disclosure. 
Fig.~\ref{fig:logging_statement_generation_examle} demonstrates Case 21. 
The extracted original code silently exits without providing feedback when users attempt to enable the \texttt{Shared} \texttt{Cache} module without setting \texttt{mapreduce.framwork.name} as \texttt{yarn}.
In contrast, the enhanced version incorporates a logging statement that explicitly links the missing parameter to component initialization failures, enabling immediate diagnosis and guiding users to apply the correct cluster resource management settings.
\answerbox{%
    \textbf{Answer to RQ1:} 
    \texttt{ConfLogger} achieves 100\% accuracy in diagnosing misconfigurations, eliminating silent failures in original implementations. Among all, 80\% cases are resolved via direct parameter extraction, significantly reducing diagnostic overhead.
}
\subsubsection{\textbf{RQ2: How does \texttt{ConfLogger} comparing other logging tools on configuration logging?}}
We compare \texttt{ConfLogger} with UniLog~\cite{xu2024unilog} and SCLogger~\cite{li2024go} on Benchmark II by measuring how many logging statements added manually can be added automatically by these tools, following previous work~\cite{ding2012ErrLog}.
\begin{table*}[tbp]
\small
  \caption{Results against baseline tools on existing logging practice.$^{*}$ Despite methodological concerns from existing logs' poor diagnosability, we include text similarity analysis for evaluation completeness. }
  \label{tab:rq2-result}
    \begin{tabular}{c||c||cccccccccc} 
    \toprule 
    \multirow{2}{*}{\textbf{Systems}} & \multirow{2}{*}{\textbf{Coverage}} &\multicolumn{2}{c||}{\textbf{Log Level}} & \multicolumn{3}{c||}{\textbf{Logging Variable}} & \multicolumn{4}{c}{\textbf{Logging Text$^{*}$}} \\ 
    {} & {} &\multirow{1}{*}{\textit{LA}} & \multicolumn{1}{c||}{\textit{AOD}} & \textit{PRECISION} & \textit{RECALL} & \multicolumn{1}{c||}{\textit{F1}} & \textit{BLEU-1} & \textit{BLEU-4}& \textit{ROUGE-1} & \textit{ROUGE-L}\\
    \toprule
    {\textit{CL}-}\texttt{{Storm}} &  {\textbf{80\%~(8/10)}} & \textbf{{0.875}} & \multicolumn{1}{c||}{\textbf{0.938}}  & {{0.630}}  & \textbf{{0.343}}  & \multicolumn{1}{c||}{{0.533}} & \textbf{{0.173}} & \textbf{{0.025}} & {0.344}  & {0.267}   \\

    \textit{UL}-\texttt{{Storm}} &  {40\%~(4/10)} & {0.750} & \multicolumn{1}{c||}{0.875}  & {/}  & {0}  & \multicolumn{1}{c||}{/} & {0.126} & \textbf{{0.025}} & \textbf{{0.358}}  & \textbf{{0.323}}   \\

    \textit{SL}-\texttt{{Storm}} &  {70\%~(7/10)} & {0} & \multicolumn{1}{c||}{0.429}  & \textbf{{0.750}}  & {0.171}  & \multicolumn{1}{c||}{\textbf{0.643}} & {0.084} & \textbf{{0.025}} & {0.196}  & {0.179}   \\
    
    \hdashline
    
    {\textit{CL}-}\texttt{{Hbase}}& \textbf{{100\%~(2/2)}} & {0} & \multicolumn{1}{c||}{0.667}  & {/}  & {0}  & \multicolumn{1}{c||}{/} & \textbf{{0.282}} & \textbf{{0.022}} & \textbf{{0.342}}  & \textbf{{0.299}}   \\
    \textit{UL}-\texttt{{Hbase}} &  \textbf{{100\%~(2/2)}} & \textbf{{1.000}} & \multicolumn{1}{c||}{\textbf{1.000}}  & {/}  & {0}  & \multicolumn{1}{c||}{/} & {0.133} & {0.020} & {0.307}  & {0.227}   \\

    \textit{SL}-\texttt{{Hbase}} &  {\textbf{100\%(2/2)}} & {\textbf{1.000}} & \multicolumn{1}{c||}{\textbf{1.000}}  & {/}  & {0}  & \multicolumn{1}{c||}{/} & {0.154} & {\textbf{0.023}} & {0.217}  & {0.130}   \\
    \hdashline

    {\textit{CL}-}\texttt{{Alluxio}}& \textbf{{100\%~(2/2)}} & \textbf{{0.500}} & \multicolumn{1}{c||}{\textbf{0.833}}  & \textbf{{1.000}}  & \textbf{{0.583}}  & \multicolumn{1}{c||}{\textbf{0.733}} & \textbf{{0.265}} & \textbf{{0.032}} & \textbf{{0.296}}  & \textbf{{0.296}}  \\
    \textit{UL}-\texttt{{Alluxio}} &  {50\%~(1/2)} & {0} & \multicolumn{1}{c||}{0.333}  & \textbf{{1.000}}  & {0.333}  & \multicolumn{1}{c||}{0.500} & {0.070} & {0.016} & {0.000}  & {0.000}   \\

    \textit{SL}-\texttt{{Alluxio}} &  \textbf{{100\%~(2/2)}} & \textbf{{0.500}} & \multicolumn{1}{c||}{0.667}  & {/}  & {0}  & \multicolumn{1}{c||}{/} & {0.035} & {0.008} & {0.105}  & {0.105}   \\
    \hdashline
    
    {\textit{CL}-}\texttt{{HCommon}}& {58\%~(7/12)} & {0.429} & \multicolumn{1}{c||}{0.738}  & {\textbf{0.694}}  & \textbf{{0.833}}  & \multicolumn{1}{c||}{\textbf{0.744}} & {0.128} & {0.018} & {0.293}  & {0.242}  \\
    \textit{UL}-\texttt{{HCommon}} &  \textbf{{75\%~(9/12)}} & \textbf{{0.778}} & \multicolumn{1}{c||}{\textbf{0.889}}  & {0.417}  & {0.312}  & \multicolumn{1}{c||}{0.389} & {\textbf{0.210}} & \textbf{{0.107}} & \textbf{{0.466}}  & \textbf{{0.426}}   \\

    \textit{SL}-\texttt{{HCommon}} &  {50\%~(6/12)} & {0.333} & \multicolumn{1}{c||}{0.694}  & {0.250}  & {0.250}  & \multicolumn{1}{c||}{0.292} & {0.144} & {0.024} & {0.288}  & {0.272}   \\
    \hdashline
    
    \textit{CL}-\texttt{{Mapreduce}}& \textbf{{81\%~(17/21)}} & \textbf{{0.824}} & \multicolumn{1}{c||}{0.882}  & {\textbf{0.590}}  & {\textbf{0.333}}  &\multicolumn{1}{c||}{\textbf{0.437}} & {\textbf{0.105}} & {0.012} & {0.200}  & {0.175}  \\
    \textit{UL}-\texttt{{Mapreduce}} &  {71\%~(15/21)} & {0.733} & \multicolumn{1}{c||}{\textbf{0.889}}  & {0.538}  & {0.250}  & \multicolumn{1}{c||}{0.397} & {0.095} & {{0.019}} & {{0.277}}  & {{0.236}}   \\

    \textit{SL}-\texttt{{Mapreduce}} &  {67\%~(14/21)} & {0.571} & \multicolumn{1}{c||}{0.762}  & {0.312}  & {0.150}  & \multicolumn{1}{c||}{0.367} & {0.102} & \textbf{{0.022}} & {\textbf{0.316}}  & \textbf{{0.297}}   \\
    \hdashline
    
    {\textit{CL}-}\texttt{{Yarn}}& \textbf{{67\%~(6/9)}} & \textbf{{0.500}} & \multicolumn{1}{c||}{\textbf{0.833}}  & \textbf{{0.667}}  & \textbf{{0.611}}  & \multicolumn{1}{c||}{\textbf{0.660}} & {{0.080}} & {{0.015}} & {0.185}  & {0.185}     \\
    \textit{UL}-\texttt{{Yarn}} &  {56\%~(5/9)} & {0.400} & \multicolumn{1}{c||}{0.767}  & \textbf{{0.667}}  & {0.267}  & \multicolumn{1}{c||}{0.500} & {0.057} & {0.010} & \textbf{{0.239}}  & \textbf{{0.226}}   \\

    \textit{SL}-\texttt{{Yarn}} &  {44\%~(4/9)} & {0.250} & \multicolumn{1}{c||}{0.750}  & {0.500}  & {0.250}  & \multicolumn{1}{c||}{0.500} & {\textbf{0.098}} & {\textbf{0.018}} & {0.179}  & {0.179}   \\
    \hdashline
    
    \textit{CL}-\texttt{{HDFS}} & \textbf{{71\%~(22/31)}} & {{0.591}} & \multicolumn{1}{c||}{{0.871}}  & {0.352}  & {\textbf{0.433}}  & \multicolumn{1}{c||}{\textbf{0.365}} & {{0.147}} & {0.020} & {{0.246}}  & {{0.194}}   \\
    \textit{UL}-\texttt{{HDFS}} &  \textbf{{71\%~(22/31)}} & {0.591} & \multicolumn{1}{c||}{0.841}  & {\textbf{0.400}}  & {0.258}  & \multicolumn{1}{c||}{0.333} & {0.121} & {{0.028}} & {0.168}  & {0.146}   \\

    \textit{SL}-\texttt{{HDFS}} &  {48\%~(15/31)} & {\textbf{0.733}} & \multicolumn{1}{c||}{\textbf{0.883}}  & {0.333}  & {0.250}  & \multicolumn{1}{c||}{0.292} & {\textbf{0.198}} & \textbf{0.046} & \textbf{0.276}  & \textbf{0.276}   \\
    \hdashline
    
    \textit{CL}-\texttt{{ZooKeeper}}& \textbf{{100\%~(3/3)}} & \textbf{{0.667}} & \multicolumn{1}{c||}{\textbf{0.778}}  & {0.750}  & {0.500}  & \multicolumn{1}{c||}{0.750} & {0.072} & {0.011} & {0.174}  & {0.174}   \\
    \textit{UL}-\texttt{{ZooKeeper}} &  {33\%~(1/3)} & {0} & \multicolumn{1}{c||}{0.333}  & {\textbf{1.000}}  & \textbf{{1.000}}  & \multicolumn{1}{c||}{\textbf{1.000}} & \textbf{{0.833}} & \textbf{{0.760}} & \textbf{{0.909}}  & \textbf{{0.909}}   \\
    \textit{SL}-\texttt{{ZooKeeper}} &  {33\%~(1/3)} & {0} & \multicolumn{1}{c||}{0.667}  & {/}  & {0}  & \multicolumn{1}{c||}{/} & {0.077} & {0.016} & {0.222}  & {0.222}   \\
    \hline
    \bottomrule 
    \textit{CL}-\textit{Average} & \textbf{{74\%~(67/90)}} & {{0.642}} & \multicolumn{1}{c||}{\textbf{0.853}}  & \textbf{{0.541}}  & {\textbf{0.459}}  & \multicolumn{1}{c||}{\textbf{0.501}} & {\textbf{0.135}} & {0.018} & {0.247}  & {0.207} \\  
    \textit{UL}-\textit{Average} &  {66\%~(59/90)} & {\textbf{0.644}} & \multicolumn{1}{c||}{0.845}  & {{0.498}}  & {0.256}  & \multicolumn{1}{c||}{0.397} & {0.134} & \textbf{{0.048}} & \textbf{{0.274}}  & {{0.244}}   \\
    \textit{SL}-\textit{Average} &  {57\%~(51/90) \ } & {0.490} & \multicolumn{1}{c||}{0.747}  & {0.357}  & {0.193}  & \multicolumn{1}{c||}{0.352} & {0.131} & {0.029} & {0.260}  & \textbf{0.247}   \\
    \bottomrule 
  \end{tabular} 
\end{table*}

\begin{figure*}[th]
    \centering
    \includegraphics[width=0.9\linewidth]{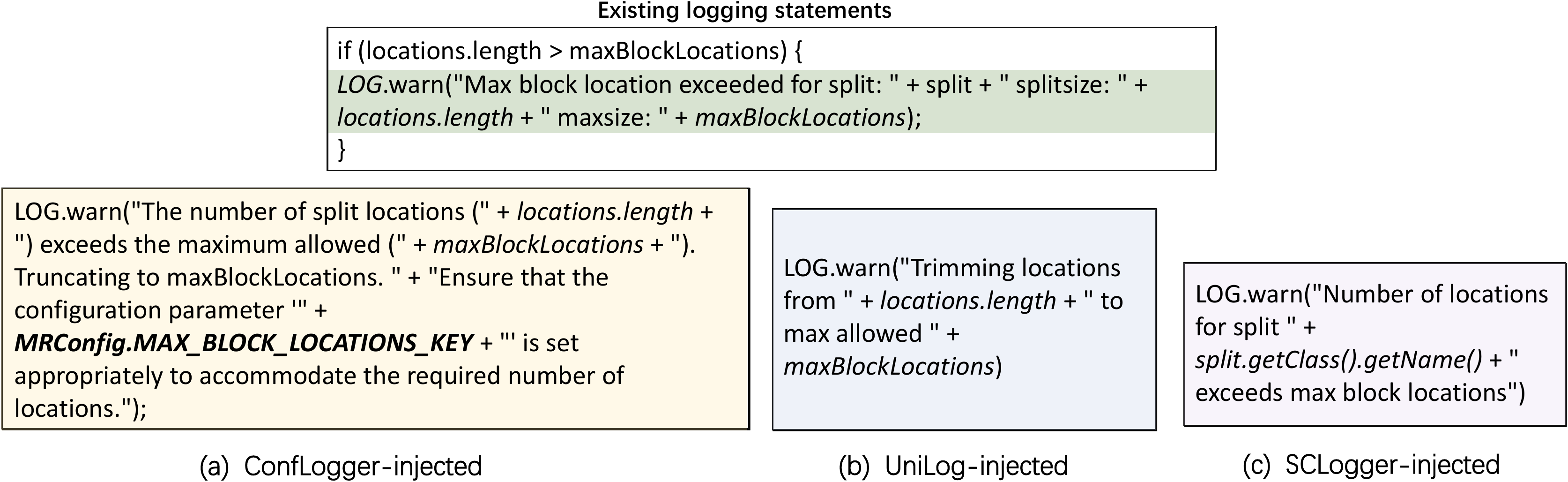}
    \caption{A case example on Benchmark II.}
    \label{fig:rq2-cases}
\end{figure*}

\textbf{\textit{Experiment Setting.}}
UniLog employs in-context learning (ICL) to retrieve top-5 similar log method examples via KNN from an external repository, prompting LLMs for target logging, while SCLogger utilizes static program analysis to extract call-chain contexts and applies BM25-based ICL for automated logging.
We reproduce a naive version of SCLogger, with the code slice graph construction and ICL strategy, under the agreement of the authors. 
These tools lack mechanisms on identification of configuration-sensitive code, as they focus on general logging automation rather than configuration-specific requirements. To enable comparative evaluation, we adapted these tools by pre-populating project-specific logging methods and manually designating target methods to meet their input constraints. 
While this adaptation introduces bias by bypassing \texttt{ConfLogger}'s automated configuration-sensitive code localization, the comparison remains valid to demonstrate its superiority in configuration scenarios over general-purpose loggers, being the first configuration-aware methodology.

\textbf{\textit{Metrics.}}
We evaluate logging quality through position accuracy (PA), log-level metrics, variable relevance, and textual similarity. PA quantifies positional validity as 1 only if the predicted log line is within one line of the ground truth and resides in the same code block; otherwise, PA is 0. System coverage is calculated as $C = \frac{N_{PA=1}}{N_{t}}$ where $N_{PA=1}$ counts valid log injections and $N_{t}$ represents the total ground truth log points.
For cases where PA equals 1, we further evaluate log-level alignment through Level Accuracy (LA), which measures exact matches between predicted and actual log levels, and Average Ordinal Distance (AOD), defined as $AOD = 1-\frac{Dist(L_T, L_I)}{maxDist(L_T)}$ where $L_T$ and $L_I$ enote the ground truth and injected log levels, respectively. Variable relevance is assessed using Precision $P=\frac{var_t \cap var_i}{var_i}$, Recall $R=\frac{var_t \cap var_i}{var_t}$, $F1= \frac{2PR}{P+R}$ (omit if no variables) where $var_t$ represents the ground truth variable and $var_i$ denotes the injected variable; 
Textual similarity is measured via BLEU-{1, 4} and ROUGE-{1, L} scores, which compute n-gram overlaps between generated and actual log texts (range: 0–1). Metrics for log levels, variables, and texts are evaluated exclusively for PA=1 cases, as incorrect log positions invalidate the purpose of downstream analysis.

\textbf{\textit{Results.}}
Results are summarized in Table~\ref{tab:rq2-result}. 
\texttt{ConfLogger} (CL) outperforms UniLog (UL) and SCLogger (SL) in diagnostic effectiveness, achieving higher coverage (74\% vs. UL:66\%, SL:57\%), log-level precision (AOD 0.853 vs. UL:0.845, SL:0.747) and optimal logging variable performance~(F1 0.501 vs. UL:0.397, SL: 0.352).
For coverage, CL’s lower coverage in certain cases, particularly within catch blocks, stems from its intentional insensitivity to generic exception-handling paths that lack explicit validation logic statements(e.g., missing if-else checking statements in such blocks).
While UL and SL suffer from coverage limitations due to their inability to prioritize configuration-sensitive code segments. 
UL’s ICL strategy struggles to emphasize critical operations in lengthy code snippets, 
while SL’s inclusion of an extended code slice graph within two hops introduces distracting context, further degrading its performance. 
Both methods tend to miss log points in configuration validation, followed by specific handling operations (e.g., reverting variables defined/passed outside the branch scope to default values), as they fail to highlight specific logging scopes for LLMs.
For log levels, CL tends to prioritize “ideal-state” monitoring (e.g., service availability, parameter compliance), leading to higher severity assignments (e.g., WARN for disabled services).
Such a design trade-off reduces its Level Accuracy (LA) compared to UL’s ICL-driven mimicry of historical patterns, however, with higher AOD. 
SL’s reliance on BM2.5 sampling for context retrieval underperforms UL’s KNN-based ICL, likely due to suboptimal similarity matching in configuration-specific scenarios.
For variable logging, CL’s advantage arises from two mechanisms: 
its ability to trace of configuration parameters and contextual emphasis on condition variables. 
UL and SL, lacking such capabilities, fail to retain critical variables in long code contexts.
Text similarity metrics reflect CL’s deliberate departure from legacy logs’ poor diagnosability. 
By injecting configuration constraints and mitigation guidance, CL prioritizes actionable diagnostics over syntactic replication.

Fig.~\ref{fig:rq2-cases} demonstrates a case that exactly explains the reasons for the performance divergence.
The original code on the upper side highlights the printed variables marked in italics. 
The lower side shows logging statements injected by the tools.
With tracked parameter identifier and specified configuration-sensitive code segments as context, CL highlights the configuration parameter\footnote{MRConfig.MAX\_BLOCK\_LOCATIONS\_KEY}~(marked in bold), whose value is carried by the condition variable\footnote{maxBlockLocations}.
Neither UL nor SL is capable of emphasizing such configuration-sensitive context to LLMs, thus they fail to log the corresponding parameter.
The original code also fails to indicate the related configuration parameter, demonstrating limited diagnosability.
UL achieves higher similarity on the logging variables and the text; however,
CL highlights the crucial configuration parameter leading to misconfiguration and includes more instructive text, showing higher diagnostic value.

\answerbox{%
    \textbf{Answer to RQ2:} 
    \texttt{ConfLogger} achieves 74\% log coverage (vs. 66\% and 57\%) and 0.853 AOD (vs. 0.845 and 0.747) with superior variable logging (F1=0.541 vs. 0.397 and 0.352), demonstrating optimal configuration-aware diagnosability.Its intentionally low text similarity prioritizes diagnostic capabilities over syntactic replication.
}
\subsubsection{\textbf{RQ3: How would the source identification strategy affect logging?}}
To assess the impact of the source identification strategy, we compare the original and an experimental variant by assessing the efficiency of labeling the configuration engines and the quality of the extracted configuration-sensitive code block.
\begin{table*}[!th]
\small
  \caption{Comparison results on source statement identification strategy versus the RN variant.}
  \label{rq3:ablation-source}
    \begin{tabular}{cccccccccc} 
    \toprule 
    \multirow{2}{*}{\textbf{Systems}} & \multirow{2}{*}{\textbf{Version}} &\multirow{2}{*}{\textbf{Strategy Time(s)}} &\multirow{2}{*}{\textbf{\# Engines}}& \multicolumn{4}{c}{\textbf{Overall Configuration Method}} & \multicolumn{2}{c}{\textbf{Valid Configuration Method}} \\
    {} & {} & {} & {}& \multirow{1}{*}{\# Valid} & \multirow{1}{*}{\# Invalid} & \multirow{1}{*}{\# Total} & \multirow{1}{*}{\% Invalid Rate} & \multirow{1}{*}{\# Specific Case}  & \multirow{1}{*}{\% Specific Rate} \\
    \toprule
   \multirow{2}{*}{ \texttt{{Storm}}} &{\texttt{RN}} &{50.338} & {2}& {55} & {68}& \textbf{{123}} & {55.30}  & \multicolumn{1}{c}{22} & {40.000}\\
    {} &{\texttt{Auto}} & {\textbf{1.465}}& {\textbf{4}}&\textbf{{65}} & \textbf{{0}} & {65}& {\textbf{0}}  & \multicolumn{1}{c}{\textbf{33}}& {\textbf{50.769}}\\
      \hdashline
     \multirow{2}{*}{\texttt{{Hbase}}} &{\texttt{RN}}&{197.209}& {2}& {17} & {0}& {17} & {\textbf{0}}  & \multicolumn{1}{c}{0} & {0}\\
    {} &{\texttt{Auto}} & {\textbf{1.810}}& {\textbf{8}} &\textbf{{19}} & {\textbf{0}} & \textbf{{19}}& {\textbf{0}}  & \multicolumn{1}{c}{\textbf{2}}& {\textbf{10.526}}\\
    \hdashline
     \multirow{2}{*}{\texttt{{Alluxio}}} &{\texttt{RN}}&{298.108}& {2}& {3} & {\textbf{0}}& {3} & {\textbf{0}}  & \multicolumn{1}{c}{1} & {33.333}\\
    {} &{\texttt{Auto}} & {\textbf{0.414}}& {\textbf{10}} &\textbf{{24}} & \textbf{{0}} & \textbf{{24}}& \textbf{{0}}  & \multicolumn{1}{c}{\textbf{23}}& {\textbf{95.833}}\\
    \hdashline
    \multirow{2}{*}{\texttt{{HCommon}}} &{\texttt{RN}}& {189.52}& {1}& \textbf{{77}} & {6} &\textbf{{83}} & {7.20}  & \multicolumn{1}{c}{\textbf{19}}& {\textbf{24.675}}\\
    {}  &{\texttt{Auto}} & {\textbf{10.353}} & \textbf{{89}} & {71} & \textbf{{5}} & {76}& {\textbf{6.60}}  & \multicolumn{1}{c}{13}& {18.310}\\
    \hdashline
    \multirow{2}{*}{\texttt{{Mapreduce}}} &{\texttt{RN}}&{71.109}& {1}& {95} & {\textbf{8}} & {103} & {7.80} & \multicolumn{1}{c}{3} & {3.158}\\
    {} &{\texttt{Auto}} & {\textbf{0.616}}& \textbf{{2}} &\textbf{{101}} & \textbf{{8}} & \textbf{{109}} & \textbf{{7.30}}  & \multicolumn{1}{c}{\textbf{10}}& {\textbf{9.901}}\\
    \hdashline
    \multirow{2}{*}{\texttt{{Yarn}}} &{\texttt{RN}}&{162.198}& {1}& \textbf{{38}} & \textbf{{1}} & \textbf{{39}} & {\textbf{2.60}}  & \multicolumn{1}{c}{\textbf{5}} & {\textbf{13.158}}\\
    {} &{\texttt{Auto}} & {\textbf{0.846}}& \textbf{{5}}&{37} & {2} & \textbf{{39}}& {5.10}  & \multicolumn{1}{c}{4}& {10.811}\\
\hdashline
    \multirow{2}{*}{\texttt{{HDFS}}} &{\texttt{RN}}&{102.409}& {1}& {68} & {3}& {71} & {{4.20}}  & \multicolumn{1}{c}{4} & {5.882}\\

    {}  &{\texttt{Auto}} & {\textbf{11.754}}& \textbf{{20}} &\textbf{{102}} & \textbf{{1}} & \textbf{{103}} & \textbf{{1.00}}  & \multicolumn{1}{c}{\textbf{40}}& {\textbf{39.216}}\\
\hdashline
    \multirow{2}{*}{\texttt{{ZooKeeper}}} &{\texttt{RN}}&{166.677}& {2}& {9} & {45}& \textbf{{54}} & {83.30}  & \multicolumn{1}{c}{0} & {0}\\
    {} &{\texttt{Auto}} & {\textbf{4.172}} & \textbf{{81}} &\textbf{{21}} & \textbf{{27}} & {48}& {\textbf{56.30}}& \multicolumn{1}{c}{\textbf{16}}& {\textbf{76.190}}\\
    \hline
    \bottomrule 
     \multirow{2}{*}{{\textit{Average}}}&{\texttt{RN}}&{154.696}& {1.5}& {45.250} & {16.375}& {\textbf{61.625}} & {26.60}  & \multicolumn{1}{c}{6.750} & {14.917}\\
    {} &{\texttt{Auto}}& {\textbf{3.929}} & {\textbf{27.375}}& {\textbf{55}} &{\textbf{5.375}} & {60.375} & {\textbf{8.90}}  & \multicolumn{1}{c}{\textbf{17.625}}& {\textbf{32.045}}\\
    \bottomrule 
  \end{tabular} \\
\end{table*}

\textbf{\textit{Experiment Setting.}}
The RN (Random-NoType) variant removes the automated strategy, instead randomly selecting one seed configuration engine from ConfLogger's configuration engine list grasped by the automated source identification strategy. 
Engineers then manually trace general configuration engines, identify getter functions for taint analysis rules. 
Conventional approaches only require locating configuration engine types regardless of specific parameters as RN implements.
We deem the original version as Auto, and compare it with the RN version.

\textbf{\textit{Metrics.}}
We introduce metrics:
(1) configuration class localization time, 
(2) total labeled configuration engines, 
(3) extracted configuration-sensitive code, and (4) valid/invalid/version-specific cases~(where version-specific cases are those extracted exclusively by a single version).
The specific rate is computed by $r = \frac{N_{s}}{N_v}$, where $N_{s}$ is the count of version-specific cases and $N_v$ is the number of valid cases.

\textbf{\textit{Results.}}
Table~\ref{rq3:ablation-source} shows the details. 
The experimental results demonstrate that \texttt{ConfLogger} (Auto version) significantly outperforms the RN version across multiple metrics. 
Strategy Time is drastically reduced in all systems with Auto, averaging 3.929s compared to RN’s 154.696s, reflecting a 97.5\% improvement in efficiency. For example, Storm’s Auto version completes in 1.465s versus RN’s 50.338s, while Alluxio’s Auto version achieves 0.414s compared to RN’s 298.108s.
Auto also exhibits superior configuration validity, with an average 8.90\% invalid rate versus RN’s 26.60\%. 
Systems like Storm and HBase using Auto achieve 0\% invalid configurations, eliminating errors entirely. 
Additionally, Auto supports more engines (average 27.375 vs. RN’s 1.5), enhancing scalability.
Notably, Auto excels in specific case tracking, achieving a 32.045\% average rate (vs. RN’s 14.917\%). 
For instance, Alluxio’s Auto version tracks 95.833\% of specific cases (vs. 33.333\% for RN), and Zookeeper’s Auto version achieves 76.190\% (vs. 0\% for RN).
RN's ability to locate more configuration-sensitive code segments because it has no restrictions on parameter types of getter methods, thus enabling more identified source statements.
However, such lack of restrictions results in more invalid code segments.
In summary, \texttt{ConfLogger} (Auto) delivers faster execution, higher accuracy, and better scalability while maintaining robustness across diverse systems.
\answerbox{%
    \textbf{Answer to RQ3:} 
    \texttt{ConfLogger} achieves a 39.36× speedup in configuration engine identification, with an 8.9\% invalid detection rate and 32\% specific success rate. These quantitative outcomes validate its scalability and precision, substantially reducing manual operational overhead in configuration automation.  
}
\begin{figure}[t]
    \centering
    \includegraphics[width=\linewidth]{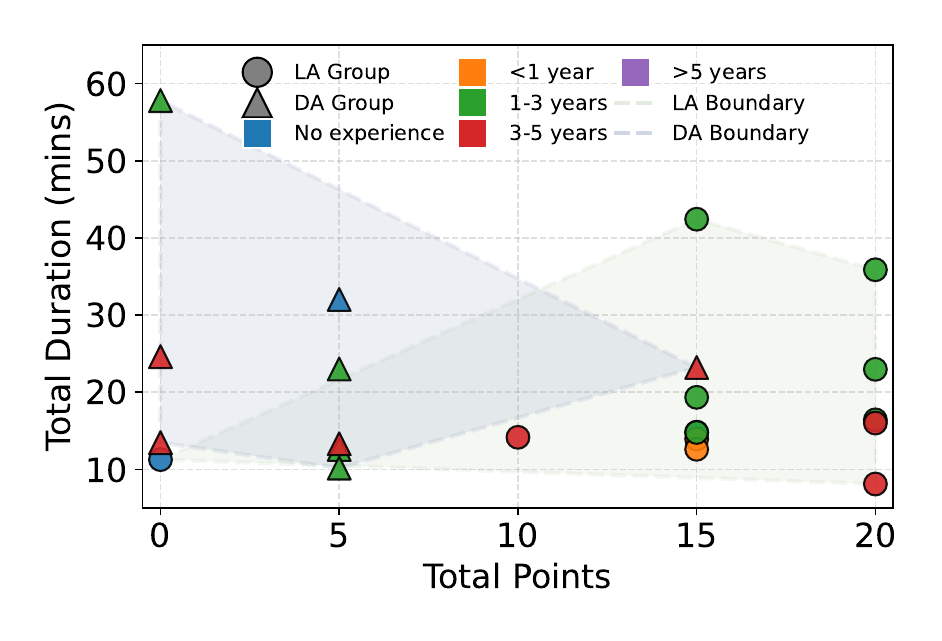}
    \vspace{-0.3in}
    \caption{Results of Group LA and DA on total points and duration.}
    \label{fig:rq3}
    \vspace{-10pt}
\end{figure}
\subsubsection{\textbf{RQ4: To what extent does ConfLogger help users in misconfiguration diagnosis? (Practical user study)}}
To evaluate \texttt{ConfLogger}'s practicality in human-centered misconfiguration diagnosis, we conduct a controlled user study.

\textit{\textbf{Experiement Setting.}}
The user study employs a between-subjects experimental design~\cite{yuan2012improving,hampton2018between} with five representative misconfiguration scenarios sampled from Benchmark I. 
Participants are randomly allocated to two experimental conditions: the Documentation-Assisted (DA) group using official configuration documents, and the Log-Assisted (LA) group utilizing \texttt{ConfLogger}'s enhanced logs, with equivalent task complexity across both groups. The timed experiment restricted task completion to 60 minutes, with quality control measures excluding responses completed in under 8 minutes from the final dataset.

\textit{\textbf{Metrics.}}
To assess diagnostic performance, we implement a scoring system awarding 5 points for correct parameter identification and 0 point for incorrect responses.
Participants' development experience is categorized into five proficiency levels (No experience, <1, 1-3, 3-5, >5 years) to contextualize performance analysis.

\textit{\textbf{Results.}}
Fig.~\ref{fig:rq3} compares diagnostic performance between LA and DA groups. LA Group data points (circles) cluster in the lower-right quadrant (green dashed boundary), demonstrating a 1.25× speedup in diagnostic time (18.68 vs. 23.36 minutes) and 251.4\% accuracy improvement (15.38 vs. 4.38 scores) over DA Group,
demonstrating efficiency and accuracy improvements.
In contrast, DA Group data points (triangles) concentrate in the upper-left quadrant (blue dashed boundary), reflecting prolonged efforts for lower scores, likely due to inefficient parameter searches in documentation. Notably, DA Group exhibits sparse high-score outliers requiring disproportionate time.
Participant experience analysis reveals LA Group’s robustness: even inexperienced users (0–1 year) achieved scores of 15–20 with minimal time, except one novice who scored zero after early abandonment. Conversely, DA Group yielded zero scores even among experienced developers (3–5 years), underscoring enhanced logging’s efficacy in reducing diagnostic effort.
Taking the demonstrated case in Fig.~\ref{fig:logging_statement_generation_examle} for example.
we simulate a failure scenario where the \texttt{Shared Cache} component in a MapReduce system fails to activate, prompting participants to identify the correct parameter among ten options. LA Group receives enhanced logs, while DA Group relies on ambiguous documentation for \texttt{mapreduce.framework.name}, which lacks explicit parameter-component correlation. Results show 85\% accuracy in LA Group (including 27\% of inexperienced users) versus 11\% in DA Group, where the sole correct response resulted from a guess.
Enhanced logs enabled diagnostic success even among inexperienced users (0–1 year), while traditional documentation led to inefficiencies (e.g., failures by experienced developers due to ambiguous parameter contexts).
\vspace{-10pt}
\answerbox{%
    \textbf{Answer to RQ4:} 
    \texttt{ConfLogger} substantially reduces diagnostic effort with enhanced logs, achieving a 251.4\% accuracy improvement and resolving issues 1.25× faster than documentation-based approaches. These enhancements also enable successful diagnosis even for inexperienced users.
}
\section{Threats to Validity}
This section systematically examines the internal and external validity of \texttt{ConfLogger}, addressing potential threats and corresponding mitigation strategies.

\textbf{Internal Threats.}
The primary internal validity concern stems from potential false positives in our program analysis. The PDG-based taint analysis, which employs control dependence for inter-procedural analysis, may lead to over-tainting phenomena~\cite{chen2020understanding,schwartz2010all}. To mitigate this, we implemented two safeguards: (1) restricting tainted path lengths to a maximum of 30 edges in our experimental settings, and (2) providing an optional control dependence deactivation mechanism during PDG construction.
An additional validity consideration arises from our validation strategy for \texttt{getter} method calls, which lacks parameter type constraints. While this optimization enhances efficiency, it may permit false positive confirmations. A proposed mitigation involves disabling confirmation procedures for such method calls.
A third internal threat is a potential for data leakage from the LLM's pre-training data. The LLM was trained on data up to October 2023, introducing a residual risk that some benchmark code, particularly from Benchmark II, may have been seen. While this could potentially inflate absolute performance, the comparative performance of our study remains valid, as all evaluated tools use the same base LLM. For Benchmark I, this risk is negligible as the log-enhanced code versions were generated by \texttt{ConfLogger}. 

\textbf{External Threats.}
Two primary factors may affect generalizability. First, our analysis framework assumes standard configuration practices and is implemented in Java. While our core methodology is language-agnostic, adapting it to other languages would require modular adjustments. Second, our bytecode-to-source code mapping mechanism necessitates the concurrent availability of both source files and compiled JARs. This dependency could be alleviated by embedding source code within JAR packages during deployment.
Another external threat arises from the impact of our approach on other diagnostic problems. While \texttt{ConfLogger} enhances logs for configuration issues, the added information could potentially introduce noise for diagnostic modules designed for other problem types. 
However, as the enriched logs help quickly exclude common configuration issues, the benefits outweigh this drawback.
Furthermore, since \texttt{ConfLogger} operates primarily within configuration-sensitive code blocks, its impact on diagnosing non-configuration issues is inherently constrained. Future work could explore strategies to balance this trade-off by selectively including other relevant variables.

\section{Conclusion} 
While modern software systems offer large configuration spaces for customization, they also put a higher requirement on diagnosing configuration errors.
This paper introduces the idea of configuration logging to expose run-time configuration details by inserting essential logging statements.
Based on this insight, we design \texttt{ConfLogger} to enhance system logging practices specifically for configurations, thereby benefiting in diagnosing relevant errors later. 
More specifically, \texttt{ConfLogger} consists of two key components: 
a configuration-sensitive code identification component realized by a taint analysis module to track configuration usage and an LLM-empowered logging statement generation component.
Our evaluations on eight mature systems confirm the effectiveness of \texttt{ConfLogger} in identifying and logging critical configuration-related information. 
Moreover, our user study underscores the practical applicability of \texttt{ConfLogger} in real-world misconfiguration diagnosis scenarios. 

\begin{acks}
 We appreciate all the anonymous reviewers for their valuable and practical comments. 
 The work  was supported by the National Key Research and Development Program of China (2023YFB2704100), the National Natural Science Foundation of China (No. 62202511), CCF - Sangfor ‘Yuanwang’ Research Fund and the
Singapore Ministry of Education (MOE) Academic Research Fund (AcRF) Tier 1 grant.
\end{acks}

\newpage
\bibliographystyle{ACM-Reference-Format}
\bibliography{sample-base}


\begin{thebibliography}{63}


\ifx \showCODEN    \undefined \def \showCODEN     #1{\unskip}     \fi
\ifx \showISBNx    \undefined \def \showISBNx     #1{\unskip}     \fi
\ifx \showISBNxiii \undefined \def \showISBNxiii  #1{\unskip}     \fi
\ifx \showISSN     \undefined \def \showISSN      #1{\unskip}     \fi
\ifx \showLCCN     \undefined \def \showLCCN      #1{\unskip}     \fi
\ifx \shownote     \undefined \def \shownote      #1{#1}          \fi
\ifx \showarticletitle \undefined \def \showarticletitle #1{#1}   \fi
\ifx \showURL      \undefined \def \showURL       {\relax}        \fi
\providecommand\bibfield[2]{#2}
\providecommand\bibinfo[2]{#2}
\providecommand\natexlab[1]{#1}
\providecommand\showeprint[2][]{arXiv:#2}

\bibitem[AlDanial(2024)]%
        {cloc}
\bibfield{author}{\bibinfo{person}{AlDanial}.} \bibinfo{year}{2024}\natexlab{}.
\newblock \bibinfo{title}{cloc}.
\newblock \bibinfo{howpublished}{\url{https://github.com/AlDanial/cloc}}.
\newblock
\newblock
\shownote{Accessed: 2025-01-20}.


\bibitem[Chen and Jiang(2017)]%
        {chen2017characterizing}
\bibfield{author}{\bibinfo{person}{Boyuan Chen} {and}
  \bibinfo{person}{Zhen~Ming Jiang}.} \bibinfo{year}{2017}\natexlab{}.
\newblock \showarticletitle{Characterizing and detecting anti-patterns in the
  logging code}. In \bibinfo{booktitle}{\emph{2017 IEEE/ACM 39th International
  Conference on Software Engineering (ICSE)}}. IEEE, \bibinfo{pages}{71--81}.
\newblock


\bibitem[Chen and Jiang(2019)]%
        {chen2019extracting}
\bibfield{author}{\bibinfo{person}{Boyuan Chen} {and}
  \bibinfo{person}{Zhen~Ming Jiang}.} \bibinfo{year}{2019}\natexlab{}.
\newblock \showarticletitle{Extracting and studying the
  Logging-Code-Issue-Introducing changes in Java-based large-scale open source
  software systems}.
\newblock \bibinfo{journal}{\emph{Empirical Software Engineering}}
  \bibinfo{volume}{24} (\bibinfo{year}{2019}), \bibinfo{pages}{2285--2322}.
\newblock


\bibitem[Chen et~al\mbox{.}(2020)]%
        {chen2020understanding}
\bibfield{author}{\bibinfo{person}{Qingrong Chen}, \bibinfo{person}{Teng Wang},
  \bibinfo{person}{Owolabi Legunsen}, \bibinfo{person}{Shanshan Li}, {and}
  \bibinfo{person}{Tianyin Xu}.} \bibinfo{year}{2020}\natexlab{}.
\newblock \showarticletitle{Understanding and discovering software
  configuration dependencies in cloud and datacenter systems}. In
  \bibinfo{booktitle}{\emph{Proceedings of the 28th ACM Joint Meeting on
  European Software Engineering Conference and Symposium on the Foundations of
  Software Engineering}}. \bibinfo{pages}{362--374}.
\newblock


\bibitem[Dong et~al\mbox{.}(2016)]%
        {dong2016orplocator}
\bibfield{author}{\bibinfo{person}{Zhen Dong}, \bibinfo{person}{Artur
  Andrzejak}, \bibinfo{person}{David Lo}, {and} \bibinfo{person}{Diego Costa}.}
  \bibinfo{year}{2016}\natexlab{}.
\newblock \showarticletitle{Orplocator: Identifying read points of
  configuration options via static analysis}. In \bibinfo{booktitle}{\emph{2016
  IEEE 27th International Symposium on Software Reliability Engineering
  (ISSRE)}}. IEEE, \bibinfo{pages}{185--195}.
\newblock


\bibitem[Du et~al\mbox{.}(2017)]%
        {du2017deeplog}
\bibfield{author}{\bibinfo{person}{Min Du}, \bibinfo{person}{Feifei Li},
  \bibinfo{person}{Guineng Zheng}, {and} \bibinfo{person}{Vivek Srikumar}.}
  \bibinfo{year}{2017}\natexlab{}.
\newblock \showarticletitle{Deeplog: Anomaly detection and diagnosis from
  system logs through deep learning}. In \bibinfo{booktitle}{\emph{Proceedings
  of the 2017 ACM SIGSAC conference on computer and communications security}}.
  \bibinfo{pages}{1285--1298}.
\newblock


\bibitem[Ferrante et~al\mbox{.}(1987)]%
        {ferrante1987program}
\bibfield{author}{\bibinfo{person}{Jeanne Ferrante}, \bibinfo{person}{Karl~J
  Ottenstein}, {and} \bibinfo{person}{Joe~D Warren}.}
  \bibinfo{year}{1987}\natexlab{}.
\newblock \showarticletitle{The program dependence graph and its use in
  optimization}.
\newblock \bibinfo{journal}{\emph{ACM Transactions on Programming Languages and
  Systems (TOPLAS)}} \bibinfo{volume}{9}, \bibinfo{number}{3}
  (\bibinfo{year}{1987}), \bibinfo{pages}{319--349}.
\newblock


\bibitem[Framework(2025a)]%
        {asm}
\bibfield{author}{\bibinfo{person}{ASM Framework}.}
  \bibinfo{year}{2025}\natexlab{a}.
\newblock \bibinfo{howpublished}{\url{https://asm.ow2.io/}}.
\newblock
\newblock
\shownote{Accessed: 2025-01-20}.


\bibitem[Framework(2025b)]%
        {wala}
\bibfield{author}{\bibinfo{person}{WALA Framework}.}
  \bibinfo{year}{2025}\natexlab{b}.
\newblock \bibinfo{howpublished}{\url{https://github.com/wala/WALA}}.
\newblock
\newblock
\shownote{Accessed: 2025-01-20}.


\bibitem[Fu et~al\mbox{.}(2024)]%
        {fu2024missconf}
\bibfield{author}{\bibinfo{person}{Ying Fu}, \bibinfo{person}{Teng Wang},
  \bibinfo{person}{Shanshan Li}, \bibinfo{person}{Jinyan Ding},
  \bibinfo{person}{Shulin Zhou}, \bibinfo{person}{Zhouyang Jia},
  \bibinfo{person}{Wang Li}, \bibinfo{person}{Yu Jiang}, {and}
  \bibinfo{person}{Xiangke Liao}.} \bibinfo{year}{2024}\natexlab{}.
\newblock \showarticletitle{MissConf: LLM-Enhanced Reproduction of
  Configuration-Triggered Bugs}. In \bibinfo{booktitle}{\emph{Proceedings of
  the 2024 IEEE/ACM 46th International Conference on Software Engineering:
  Companion Proceedings}}. \bibinfo{pages}{484--495}.
\newblock


\bibitem[GPT-4o({[n.\,d.]})]%
        {gpt-4o}
\bibfield{author}{\bibinfo{person}{GPT-4o}.}
  \bibinfo{year}{[n.\,d.]}\natexlab{}.
\newblock \bibinfo{title}{GPT-4o}.
\newblock
  \bibinfo{howpublished}{\url{https://platform.openai.com/docs/models/gpt-4o}}.
\newblock
\newblock
\shownote{Accessed: 2025-01-20}.


\bibitem[Hampton(2018)]%
        {hampton2018between}
\bibfield{author}{\bibinfo{person}{James Hampton}.}
  \bibinfo{year}{2018}\natexlab{}.
\newblock \showarticletitle{The between-subjects experiment}.
\newblock In \bibinfo{booktitle}{\emph{Laboratory psychology}}.
  \bibinfo{publisher}{Psychology Press}, \bibinfo{pages}{15--37}.
\newblock


\bibitem[He et~al\mbox{.}(2017)]%
        {he2017drain}
\bibfield{author}{\bibinfo{person}{Pinjia He}, \bibinfo{person}{Jieming Zhu},
  \bibinfo{person}{Zibin Zheng}, {and} \bibinfo{person}{Michael~R Lyu}.}
  \bibinfo{year}{2017}\natexlab{}.
\newblock \showarticletitle{Drain: An online log parsing approach with fixed
  depth tree}. In \bibinfo{booktitle}{\emph{2017 IEEE international conference
  on web services (ICWS)}}. IEEE, \bibinfo{pages}{33--40}.
\newblock


\bibitem[He et~al\mbox{.}(2021)]%
        {he2021survey}
\bibfield{author}{\bibinfo{person}{Shilin He}, \bibinfo{person}{Pinjia He},
  \bibinfo{person}{Zhuangbin Chen}, \bibinfo{person}{Tianyi Yang},
  \bibinfo{person}{Yuxin Su}, {and} \bibinfo{person}{Michael~R Lyu}.}
  \bibinfo{year}{2021}\natexlab{}.
\newblock \showarticletitle{A survey on automated log analysis for reliability
  engineering}.
\newblock \bibinfo{journal}{\emph{ACM computing surveys (CSUR)}}
  \bibinfo{volume}{54}, \bibinfo{number}{6} (\bibinfo{year}{2021}),
  \bibinfo{pages}{1--37}.
\newblock


\bibitem[Huo et~al\mbox{.}(2023a)]%
        {huo2023evlog}
\bibfield{author}{\bibinfo{person}{Yintong Huo}, \bibinfo{person}{Cheryl Lee},
  \bibinfo{person}{Yuxin Su}, \bibinfo{person}{Shiwen Shan},
  \bibinfo{person}{Jinyang Liu}, {and} \bibinfo{person}{Michael~R Lyu}.}
  \bibinfo{year}{2023}\natexlab{a}.
\newblock \showarticletitle{EvLog: Identifying Anomalous Logs over Software
  Evolution}. In \bibinfo{booktitle}{\emph{2023 IEEE 34th International
  Symposium on Software Reliability Engineering (ISSRE)}}. IEEE,
  \bibinfo{pages}{391--402}.
\newblock


\bibitem[Huo et~al\mbox{.}(2023b)]%
        {huo2023semparser}
\bibfield{author}{\bibinfo{person}{Yintong Huo}, \bibinfo{person}{Yuxin Su},
  \bibinfo{person}{Cheryl Lee}, {and} \bibinfo{person}{Michael~R Lyu}.}
  \bibinfo{year}{2023}\natexlab{b}.
\newblock \showarticletitle{Semparser: A semantic parser for log analytics}. In
  \bibinfo{booktitle}{\emph{2023 IEEE/ACM 45th International Conference on
  Software Engineering (ICSE)}}. IEEE, \bibinfo{pages}{881--893}.
\newblock


\bibitem[Ji et~al\mbox{.}(2023)]%
        {ji2023survey}
\bibfield{author}{\bibinfo{person}{Ziwei Ji}, \bibinfo{person}{Nayeon Lee},
  \bibinfo{person}{Rita Frieske}, \bibinfo{person}{Tiezheng Yu},
  \bibinfo{person}{Dan Su}, \bibinfo{person}{Yan Xu}, \bibinfo{person}{Etsuko
  Ishii}, \bibinfo{person}{Ye~Jin Bang}, \bibinfo{person}{Andrea Madotto},
  {and} \bibinfo{person}{Pascale Fung}.} \bibinfo{year}{2023}\natexlab{}.
\newblock \showarticletitle{Survey of hallucination in natural language
  generation}.
\newblock \bibinfo{journal}{\emph{Comput. Surveys}} \bibinfo{volume}{55},
  \bibinfo{number}{12} (\bibinfo{year}{2023}), \bibinfo{pages}{1--38}.
\newblock


\bibitem[Jia et~al\mbox{.}(2018)]%
        {jia2018smartlog}
\bibfield{author}{\bibinfo{person}{Zhouyang Jia}, \bibinfo{person}{Shanshan
  Li}, \bibinfo{person}{Xiaodong Liu}, \bibinfo{person}{Xiangke Liao}, {and}
  \bibinfo{person}{Yunhuai Liu}.} \bibinfo{year}{2018}\natexlab{}.
\newblock \showarticletitle{SMARTLOG: Place error log statement by deep
  understanding of log intention}. In \bibinfo{booktitle}{\emph{2018 IEEE 25th
  International Conference on Software Analysis, Evolution and Reengineering
  (SANER)}}. IEEE, \bibinfo{pages}{61--71}.
\newblock


\bibitem[Jiang et~al\mbox{.}(2024)]%
        {jiang2024lilac}
\bibfield{author}{\bibinfo{person}{Zhihan Jiang}, \bibinfo{person}{Jinyang
  Liu}, \bibinfo{person}{Zhuangbin Chen}, \bibinfo{person}{Yichen Li},
  \bibinfo{person}{Junjie Huang}, \bibinfo{person}{Yintong Huo},
  \bibinfo{person}{Pinjia He}, \bibinfo{person}{Jiazhen Gu}, {and}
  \bibinfo{person}{Michael~R Lyu}.} \bibinfo{year}{2024}\natexlab{}.
\newblock \showarticletitle{LILAC: Log parsing using LLMs with adaptive parsing
  cache}.
\newblock \bibinfo{journal}{\emph{Proceedings of the ACM on Software
  Engineering}} \bibinfo{volume}{1}, \bibinfo{number}{FSE}
  (\bibinfo{year}{2024}), \bibinfo{pages}{137--160}.
\newblock


\bibitem[Keller et~al\mbox{.}(2008)]%
        {keller2008conferr}
\bibfield{author}{\bibinfo{person}{Lorenzo Keller}, \bibinfo{person}{Prasang
  Upadhyaya}, {and} \bibinfo{person}{George Candea}.}
  \bibinfo{year}{2008}\natexlab{}.
\newblock \showarticletitle{ConfErr: A tool for assessing resilience to human
  configuration errors}. In \bibinfo{booktitle}{\emph{2008 IEEE International
  Conference on Dependable Systems and Networks With FTCS and DCC (DSN)}}.
  IEEE, \bibinfo{pages}{157--166}.
\newblock


\bibitem[Li et~al\mbox{.}(2020b)]%
        {li2020qualitative}
\bibfield{author}{\bibinfo{person}{Heng Li}, \bibinfo{person}{Weiyi Shang},
  \bibinfo{person}{Bram Adams}, \bibinfo{person}{Mohammed Sayagh}, {and}
  \bibinfo{person}{Ahmed~E Hassan}.} \bibinfo{year}{2020}\natexlab{b}.
\newblock \showarticletitle{A qualitative study of the benefits and costs of
  logging from developers’ perspectives}.
\newblock \bibinfo{journal}{\emph{IEEE Transactions on Software Engineering}}
  \bibinfo{volume}{47}, \bibinfo{number}{12} (\bibinfo{year}{2020}),
  \bibinfo{pages}{2858--2873}.
\newblock


\bibitem[Li et~al\mbox{.}(2017b)]%
        {li2017log}
\bibfield{author}{\bibinfo{person}{Heng Li}, \bibinfo{person}{Weiyi Shang},
  {and} \bibinfo{person}{Ahmed~E Hassan}.} \bibinfo{year}{2017}\natexlab{b}.
\newblock \showarticletitle{Which log level should developers choose for a new
  logging statement?}
\newblock \bibinfo{journal}{\emph{Empirical Software Engineering}}
  \bibinfo{volume}{22} (\bibinfo{year}{2017}), \bibinfo{pages}{1684--1716}.
\newblock


\bibitem[Li et~al\mbox{.}(2024b)]%
        {li2024ecfuzz}
\bibfield{author}{\bibinfo{person}{Junqiang Li}, \bibinfo{person}{Senyi Li},
  \bibinfo{person}{Keyao Li}, \bibinfo{person}{Falin Luo},
  \bibinfo{person}{Hongfang Yu}, \bibinfo{person}{Shanshan Li}, {and}
  \bibinfo{person}{Xiang Li}.} \bibinfo{year}{2024}\natexlab{b}.
\newblock \showarticletitle{ECFuzz: Effective Configuration Fuzzing for
  Large-Scale Systems}. In \bibinfo{booktitle}{\emph{Proceedings of the 46th
  IEEE/ACM International Conference on Software Engineering}}.
  \bibinfo{pages}{1--12}.
\newblock


\bibitem[Li et~al\mbox{.}(2018)]%
        {li2018confvd}
\bibfield{author}{\bibinfo{person}{Shanshan Li}, \bibinfo{person}{Wang Li},
  \bibinfo{person}{Xiangke Liao}, \bibinfo{person}{Shaoliang Peng},
  \bibinfo{person}{Shulin Zhou}, \bibinfo{person}{Zhouyang Jia}, {and}
  \bibinfo{person}{Teng Wang}.} \bibinfo{year}{2018}\natexlab{}.
\newblock \showarticletitle{Confvd: System reactions analysis and evaluation
  through misconfiguration injection}.
\newblock \bibinfo{journal}{\emph{IEEE Transactions on Reliability}}
  \bibinfo{volume}{67}, \bibinfo{number}{4} (\bibinfo{year}{2018}),
  \bibinfo{pages}{1393--1405}.
\newblock


\bibitem[Li et~al\mbox{.}(2021a)]%
        {li2021challenges}
\bibfield{author}{\bibinfo{person}{Wang Li}, \bibinfo{person}{Zhouyang Jia},
  \bibinfo{person}{Shanshan Li}, \bibinfo{person}{Yuanliang Zhang},
  \bibinfo{person}{Teng Wang}, \bibinfo{person}{Erci Xu}, \bibinfo{person}{Ji
  Wang}, {and} \bibinfo{person}{Xiangke Liao}.}
  \bibinfo{year}{2021}\natexlab{a}.
\newblock \showarticletitle{Challenges and opportunities: an in-depth empirical
  study on configuration error injection testing}. In
  \bibinfo{booktitle}{\emph{Proceedings of the 30th ACM SIGSOFT International
  Symposium on Software Testing and Analysis}}. \bibinfo{pages}{478--490}.
\newblock


\bibitem[Li et~al\mbox{.}(2017a)]%
        {li2017conftest}
\bibfield{author}{\bibinfo{person}{Wang Li}, \bibinfo{person}{Shanshan Li},
  \bibinfo{person}{Xiangke Liao}, \bibinfo{person}{Xiangyang Xu},
  \bibinfo{person}{Shulin Zhou}, {and} \bibinfo{person}{Zhouyang Jia}.}
  \bibinfo{year}{2017}\natexlab{a}.
\newblock \showarticletitle{Conftest: Generating comprehensive misconfiguration
  for system reaction ability evaluation}. In
  \bibinfo{booktitle}{\emph{Proceedings of the 21st International Conference on
  Evaluation and Assessment in Software Engineering}}. \bibinfo{pages}{88--97}.
\newblock


\bibitem[Li et~al\mbox{.}(2024a)]%
        {li2024go}
\bibfield{author}{\bibinfo{person}{Yichen Li}, \bibinfo{person}{Yintong Huo},
  \bibinfo{person}{Renyi Zhong}, \bibinfo{person}{Zhihan Jiang},
  \bibinfo{person}{Jinyang Liu}, \bibinfo{person}{Junjie Huang},
  \bibinfo{person}{Jiazhen Gu}, \bibinfo{person}{Pinjia He}, {and}
  \bibinfo{person}{Michael~R Lyu}.} \bibinfo{year}{2024}\natexlab{a}.
\newblock \showarticletitle{Go static: Contextualized logging statement
  generation}.
\newblock \bibinfo{journal}{\emph{Proceedings of the ACM on Software
  Engineering}} \bibinfo{volume}{1}, \bibinfo{number}{FSE}
  (\bibinfo{year}{2024}), \bibinfo{pages}{609--630}.
\newblock


\bibitem[Li(2020)]%
        {li2020towards}
\bibfield{author}{\bibinfo{person}{Zhenhao Li}.}
  \bibinfo{year}{2020}\natexlab{}.
\newblock \showarticletitle{Towards providing automated supports to developers
  on writing logging statements}. In \bibinfo{booktitle}{\emph{Proceedings of
  the ACM/IEEE 42nd International Conference on Software Engineering: Companion
  Proceedings}}. \bibinfo{pages}{198--201}.
\newblock


\bibitem[Li et~al\mbox{.}(2020a)]%
        {li2020shall}
\bibfield{author}{\bibinfo{person}{Zhenhao Li}, \bibinfo{person}{Tse-Hsun
  Chen}, {and} \bibinfo{person}{Weiyi Shang}.}
  \bibinfo{year}{2020}\natexlab{a}.
\newblock \showarticletitle{Where shall we log? studying and suggesting logging
  locations in code blocks}. In \bibinfo{booktitle}{\emph{Proceedings of the
  35th IEEE/ACM International Conference on Automated Software Engineering}}.
  \bibinfo{pages}{361--372}.
\newblock


\bibitem[Li et~al\mbox{.}(2021b)]%
        {li2021deeplv}
\bibfield{author}{\bibinfo{person}{Zhenhao Li}, \bibinfo{person}{Heng Li},
  \bibinfo{person}{Tse-Hsun Chen}, {and} \bibinfo{person}{Weiyi Shang}.}
  \bibinfo{year}{2021}\natexlab{b}.
\newblock \showarticletitle{Deeplv: Suggesting log levels using ordinal based
  neural networks}. In \bibinfo{booktitle}{\emph{2021 IEEE/ACM 43rd
  International Conference on Software Engineering (ICSE)}}. IEEE,
  \bibinfo{pages}{1461--1472}.
\newblock


\bibitem[Lian et~al\mbox{.}(2024)]%
        {lian2024large}
\bibfield{author}{\bibinfo{person}{Xinyu Lian}, \bibinfo{person}{Yinfang Chen},
  \bibinfo{person}{Runxiang Cheng}, \bibinfo{person}{Jie Huang},
  \bibinfo{person}{Parth Thakkar}, \bibinfo{person}{Minjia Zhang}, {and}
  \bibinfo{person}{Tianyin Xu}.} \bibinfo{year}{2024}\natexlab{}.
\newblock \showarticletitle{Large Language Models as Configuration Validators}.
  In \bibinfo{booktitle}{\emph{2025 IEEE/ACM 47th International Conference on
  Software Engineering (ICSE)}}. IEEE Computer Society,
  \bibinfo{pages}{204--216}.
\newblock


\bibitem[Liao et~al\mbox{.}(2018)]%
        {liao2018you}
\bibfield{author}{\bibinfo{person}{Xiangke Liao}, \bibinfo{person}{Shulin
  Zhou}, \bibinfo{person}{Shanshan Li}, \bibinfo{person}{Zhouyang Jia},
  \bibinfo{person}{Xiaodong Liu}, {and} \bibinfo{person}{Haochen He}.}
  \bibinfo{year}{2018}\natexlab{}.
\newblock \showarticletitle{Do you really know how to configure your software?
  configuration constraints in source code may help}.
\newblock \bibinfo{journal}{\emph{IEEE Transactions on Reliability}}
  \bibinfo{volume}{67}, \bibinfo{number}{3} (\bibinfo{year}{2018}),
  \bibinfo{pages}{832--846}.
\newblock


\bibitem[Liu et~al\mbox{.}(2022)]%
        {liu2022tell}
\bibfield{author}{\bibinfo{person}{Jiahao Liu}, \bibinfo{person}{Jun Zeng},
  \bibinfo{person}{Xiang Wang}, \bibinfo{person}{Kaihang Ji}, {and}
  \bibinfo{person}{Zhenkai Liang}.} \bibinfo{year}{2022}\natexlab{}.
\newblock \showarticletitle{Tell: log level suggestions via modeling
  multi-level code block information}. In \bibinfo{booktitle}{\emph{Proceedings
  of the 31st ACM SIGSOFT International Symposium on Software Testing and
  Analysis}}. \bibinfo{pages}{27--38}.
\newblock


\bibitem[Mastropaolo et~al\mbox{.}(2022)]%
        {mastropaolo2022using}
\bibfield{author}{\bibinfo{person}{Antonio Mastropaolo}, \bibinfo{person}{Luca
  Pascarella}, {and} \bibinfo{person}{Gabriele Bavota}.}
  \bibinfo{year}{2022}\natexlab{}.
\newblock \showarticletitle{Using deep learning to generate complete log
  statements}. In \bibinfo{booktitle}{\emph{Proceedings of the 44th
  International Conference on Software Engineering}}.
  \bibinfo{pages}{2279--2290}.
\newblock


\bibitem[Rahman and Kundu(2024)]%
        {rahman2024code}
\bibfield{author}{\bibinfo{person}{Mirza~Masfiqur Rahman} {and}
  \bibinfo{person}{Ashish Kundu}.} \bibinfo{year}{2024}\natexlab{}.
\newblock \showarticletitle{Code Hallucination}.
\newblock \bibinfo{journal}{\emph{arXiv preprint arXiv:2407.04831}}
  (\bibinfo{year}{2024}).
\newblock


\bibitem[Repository(2024)]%
        {maven}
\bibfield{author}{\bibinfo{person}{Maven Repository}.}
  \bibinfo{year}{2024}\natexlab{}.
\newblock \bibinfo{title}{Maven Repository}.
\newblock \bibinfo{howpublished}{\url{https://mvnrepository.com/}}.
\newblock
\newblock
\shownote{Accessed: 2025-01-20}.


\bibitem[Schwartz et~al\mbox{.}(2010)]%
        {schwartz2010all}
\bibfield{author}{\bibinfo{person}{Edward~J Schwartz},
  \bibinfo{person}{Thanassis Avgerinos}, {and} \bibinfo{person}{David
  Brumley}.} \bibinfo{year}{2010}\natexlab{}.
\newblock \showarticletitle{All you ever wanted to know about dynamic taint
  analysis and forward symbolic execution (but might have been afraid to ask)}.
  In \bibinfo{booktitle}{\emph{2010 IEEE symposium on Security and privacy}}.
  IEEE, \bibinfo{pages}{317--331}.
\newblock


\bibitem[Shan et~al\mbox{.}(2024)]%
        {shan2024face}
\bibfield{author}{\bibinfo{person}{Shiwen Shan}, \bibinfo{person}{Yintong Huo},
  \bibinfo{person}{Yuxin Su}, \bibinfo{person}{Yichen Li}, \bibinfo{person}{Dan
  Li}, {and} \bibinfo{person}{Zibin Zheng}.} \bibinfo{year}{2024}\natexlab{}.
\newblock \showarticletitle{Face it yourselves: An llm-based two-stage strategy
  to localize configuration errors via logs}. In
  \bibinfo{booktitle}{\emph{Proceedings of the 33rd ACM SIGSOFT International
  Symposium on Software Testing and Analysis}}. \bibinfo{pages}{13--25}.
\newblock


\bibitem[SLF4J(2024)]%
        {slf4j}
\bibfield{author}{\bibinfo{person}{SLF4J}.} \bibinfo{year}{2024}\natexlab{}.
\newblock \bibinfo{title}{SLF4J}.
\newblock \bibinfo{howpublished}{\url{https://slf4j.org/}}.
\newblock
\newblock
\shownote{Accessed: 2025-01-20}.


\bibitem[Sun et~al\mbox{.}(2020)]%
        {sun2020testing}
\bibfield{author}{\bibinfo{person}{Xudong Sun}, \bibinfo{person}{Runxiang
  Cheng}, \bibinfo{person}{Jianyan Chen}, \bibinfo{person}{Elaine Ang},
  \bibinfo{person}{Owolabi Legunsen}, {and} \bibinfo{person}{Tianyin Xu}.}
  \bibinfo{year}{2020}\natexlab{}.
\newblock \showarticletitle{Testing configuration changes in context to prevent
  production failures}. In \bibinfo{booktitle}{\emph{14th USENIX Symposium on
  Operating Systems Design and Implementation (OSDI 20)}}.
  \bibinfo{pages}{735--751}.
\newblock


\bibitem[Tang et~al\mbox{.}(2015)]%
        {tang2015holistic}
\bibfield{author}{\bibinfo{person}{Chunqiang Tang}, \bibinfo{person}{Thawan
  Kooburat}, \bibinfo{person}{Pradeep Venkatachalam}, \bibinfo{person}{Akshay
  Chander}, \bibinfo{person}{Zhe Wen}, \bibinfo{person}{Aravind Narayanan},
  \bibinfo{person}{Patrick Dowell}, {and} \bibinfo{person}{Robert Karl}.}
  \bibinfo{year}{2015}\natexlab{}.
\newblock \showarticletitle{Holistic configuration management at facebook}. In
  \bibinfo{booktitle}{\emph{Proceedings of the 25th symposium on operating
  systems principles}}. \bibinfo{pages}{328--343}.
\newblock


\bibitem[Vevera et~al\mbox{.}(2025)]%
        {vevera2025cybersecurity}
\bibfield{author}{\bibinfo{person}{Adrian-Victor Vevera},
  \bibinfo{person}{Andreea~C{\u{a}}t{\u{a}}lina CR{\u{A}}CIUN},
  \bibinfo{person}{Mihail DUMITRACHE}, \bibinfo{person}{Ionut SANDU},
  \bibinfo{person}{Carmen-Ionela ROTUN{\u{A}}}, {and}
  \bibinfo{person}{Radu~Alexandru BOSTAN}.} \bibinfo{year}{2025}\natexlab{}.
\newblock \showarticletitle{Cybersecurity Challenges in Managing Domain Names.
  From DNS to ENS in the Web3 Era}.
\newblock \bibinfo{journal}{\emph{Romanian Cyber Security Journal}}
  \bibinfo{volume}{7}, \bibinfo{number}{1} (\bibinfo{year}{2025}),
  \bibinfo{pages}{97--112}.
\newblock


\bibitem[Wang et~al\mbox{.}(2023a)]%
        {wang2023conftainter}
\bibfield{author}{\bibinfo{person}{Teng Wang}, \bibinfo{person}{Haochen He},
  \bibinfo{person}{Xiaodong Liu}, \bibinfo{person}{Shanshan Li},
  \bibinfo{person}{Zhouyang Jia}, \bibinfo{person}{Yu Jiang},
  \bibinfo{person}{Qing Liao}, {and} \bibinfo{person}{Wang Li}.}
  \bibinfo{year}{2023}\natexlab{a}.
\newblock \showarticletitle{Conftainter: Static taint analysis for
  configuration options}. In \bibinfo{booktitle}{\emph{2023 38th IEEE/ACM
  International Conference on Automated Software Engineering (ASE)}}. IEEE,
  \bibinfo{pages}{1640--1651}.
\newblock


\bibitem[Wang et~al\mbox{.}(2023b)]%
        {wang2023understanding}
\bibfield{author}{\bibinfo{person}{Teng Wang}, \bibinfo{person}{Zhouyang Jia},
  \bibinfo{person}{Shanshan Li}, \bibinfo{person}{Si Zheng},
  \bibinfo{person}{Yue Yu}, \bibinfo{person}{Erci Xu},
  \bibinfo{person}{Shaoliang Peng}, {and} \bibinfo{person}{Xiangke Liao}.}
  \bibinfo{year}{2023}\natexlab{b}.
\newblock \showarticletitle{Understanding and detecting on-the-fly
  configuration bugs}. In \bibinfo{booktitle}{\emph{2023 IEEE/ACM 45th
  International Conference on Software Engineering (ICSE)}}. IEEE,
  \bibinfo{pages}{628--639}.
\newblock


\bibitem[Wang et~al\mbox{.}(2018)]%
        {wang2018misconfdoctor}
\bibfield{author}{\bibinfo{person}{Teng Wang}, \bibinfo{person}{Xiaodong Liu},
  \bibinfo{person}{Shanshan Li}, \bibinfo{person}{Xiangke Liao},
  \bibinfo{person}{Wang Li}, {and} \bibinfo{person}{Qing Liao}.}
  \bibinfo{year}{2018}\natexlab{}.
\newblock \showarticletitle{MisconfDoctor: diagnosing misconfiguration via
  log-based configuration testing}. In \bibinfo{booktitle}{\emph{2018 IEEE
  International Conference on Software Quality, Reliability and Security
  (QRS)}}. IEEE, \bibinfo{pages}{1--12}.
\newblock


\bibitem[Wei et~al\mbox{.}(2022)]%
        {wei2022chain}
\bibfield{author}{\bibinfo{person}{Jason Wei}, \bibinfo{person}{Xuezhi Wang},
  \bibinfo{person}{Dale Schuurmans}, \bibinfo{person}{Maarten Bosma},
  \bibinfo{person}{Fei Xia}, \bibinfo{person}{Ed Chi}, \bibinfo{person}{Quoc~V
  Le}, \bibinfo{person}{Denny Zhou}, {et~al\mbox{.}}}
  \bibinfo{year}{2022}\natexlab{}.
\newblock \showarticletitle{Chain-of-thought prompting elicits reasoning in
  large language models}.
\newblock \bibinfo{journal}{\emph{Advances in neural information processing
  systems}}  \bibinfo{volume}{35} (\bibinfo{year}{2022}),
  \bibinfo{pages}{24824--24837}.
\newblock


\bibitem[Xu et~al\mbox{.}(2024)]%
        {xu2024unilog}
\bibfield{author}{\bibinfo{person}{Junjielong Xu}, \bibinfo{person}{Ziang Cui},
  \bibinfo{person}{Yuan Zhao}, \bibinfo{person}{Xu Zhang},
  \bibinfo{person}{Shilin He}, \bibinfo{person}{Pinjia He},
  \bibinfo{person}{Liqun Li}, \bibinfo{person}{Yu Kang},
  \bibinfo{person}{Qingwei Lin}, \bibinfo{person}{Yingnong Dang},
  {et~al\mbox{.}}} \bibinfo{year}{2024}\natexlab{}.
\newblock \showarticletitle{UniLog: Automatic Logging via LLM and In-Context
  Learning}. In \bibinfo{booktitle}{\emph{Proceedings of the 46th IEEE/ACM
  International Conference on Software Engineering}}. \bibinfo{pages}{1--12}.
\newblock


\bibitem[Xu et~al\mbox{.}(2015)]%
        {xu2015hey}
\bibfield{author}{\bibinfo{person}{Tianyin Xu}, \bibinfo{person}{Long Jin},
  \bibinfo{person}{Xuepeng Fan}, \bibinfo{person}{Yuanyuan Zhou},
  \bibinfo{person}{Shankar Pasupathy}, {and} \bibinfo{person}{Rukma
  Talwadker}.} \bibinfo{year}{2015}\natexlab{}.
\newblock \showarticletitle{Hey, you have given me too many knobs!:
  Understanding and dealing with over-designed configuration in system
  software}. In \bibinfo{booktitle}{\emph{Proceedings of the 2015 10th Joint
  Meeting on Foundations of Software Engineering}}. \bibinfo{pages}{307--319}.
\newblock


\bibitem[Xu et~al\mbox{.}(2016)]%
        {xu2016early}
\bibfield{author}{\bibinfo{person}{Tianyin Xu}, \bibinfo{person}{Xinxin Jin},
  \bibinfo{person}{Peng Huang}, \bibinfo{person}{Yuanyuan Zhou},
  \bibinfo{person}{Shan Lu}, \bibinfo{person}{Long Jin}, {and}
  \bibinfo{person}{Shankar Pasupathy}.} \bibinfo{year}{2016}\natexlab{}.
\newblock \showarticletitle{Early detection of configuration errors to reduce
  failure damage}. In \bibinfo{booktitle}{\emph{12th USENIX Symposium on
  Operating Systems Design and Implementation (OSDI 16)}}.
  \bibinfo{pages}{619--634}.
\newblock


\bibitem[Xu and Legunsen(2019)]%
        {xu2019configuration}
\bibfield{author}{\bibinfo{person}{Tianyin Xu} {and} \bibinfo{person}{Owolabi
  Legunsen}.} \bibinfo{year}{2019}\natexlab{}.
\newblock \showarticletitle{Configuration testing: Testing configuration values
  as code and with code}.
\newblock \bibinfo{journal}{\emph{arXiv preprint arXiv:1905.12195}}
  (\bibinfo{year}{2019}).
\newblock


\bibitem[Xu et~al\mbox{.}(2013)]%
        {xu2013not}
\bibfield{author}{\bibinfo{person}{Tianyin Xu}, \bibinfo{person}{Jiaqi Zhang},
  \bibinfo{person}{Peng Huang}, \bibinfo{person}{Jing Zheng},
  \bibinfo{person}{Tianwei Sheng}, \bibinfo{person}{Ding Yuan},
  \bibinfo{person}{Yuanyuan Zhou}, {and} \bibinfo{person}{Shankar Pasupathy}.}
  \bibinfo{year}{2013}\natexlab{}.
\newblock \showarticletitle{Do not blame users for misconfigurations}. In
  \bibinfo{booktitle}{\emph{Proceedings of the Twenty-Fourth ACM Symposium on
  Operating Systems Principles}}. \bibinfo{pages}{244--259}.
\newblock


\bibitem[Yao et~al\mbox{.}(2018)]%
        {yao2018log4perf}
\bibfield{author}{\bibinfo{person}{Kundi Yao}, \bibinfo{person}{Guilherme B.~de
  P{\'a}dua}, \bibinfo{person}{Weiyi Shang}, \bibinfo{person}{Steve Sporea},
  \bibinfo{person}{Andrei Toma}, {and} \bibinfo{person}{Sarah Sajedi}.}
  \bibinfo{year}{2018}\natexlab{}.
\newblock \showarticletitle{Log4perf: Suggesting logging locations for
  web-based systems' performance monitoring}. In
  \bibinfo{booktitle}{\emph{Proceedings of the 2018 ACM/SPEC International
  Conference on Performance Engineering}}. \bibinfo{pages}{127--138}.
\newblock


\bibitem[Yin et~al\mbox{.}(2011)]%
        {yin2011empirical}
\bibfield{author}{\bibinfo{person}{Zuoning Yin}, \bibinfo{person}{Xiao Ma},
  \bibinfo{person}{Jing Zheng}, \bibinfo{person}{Yuanyuan Zhou},
  \bibinfo{person}{Lakshmi~N Bairavasundaram}, {and} \bibinfo{person}{Shankar
  Pasupathy}.} \bibinfo{year}{2011}\natexlab{}.
\newblock \showarticletitle{An empirical study on configuration errors in
  commercial and open source systems}. In \bibinfo{booktitle}{\emph{Proceedings
  of the Twenty-Third ACM Symposium on Operating Systems Principles}}.
  \bibinfo{pages}{159--172}.
\newblock


\bibitem[Yuan et~al\mbox{.}(2012a)]%
        {ding2012ErrLog}
\bibfield{author}{\bibinfo{person}{Ding Yuan}, \bibinfo{person}{Soyeon Park},
  \bibinfo{person}{Peng Huang}, \bibinfo{person}{Yang Liu},
  \bibinfo{person}{Michael~M. Lee}, \bibinfo{person}{Xiaoming Tang},
  \bibinfo{person}{Yuanyuan Zhou}, {and} \bibinfo{person}{Stefan Savage}.}
  \bibinfo{year}{2012}\natexlab{a}.
\newblock \showarticletitle{Be Conservative: Enhancing Failure Diagnosis with
  Proactive Logging}. In \bibinfo{booktitle}{\emph{10th USENIX Symposium on
  Operating Systems Design and Implementation (OSDI 12)}}.
  \bibinfo{publisher}{USENIX Association}, \bibinfo{address}{Hollywood, CA},
  \bibinfo{pages}{293--306}.
\newblock
\showISBNx{978-1-931971-96-6}
\urldef\tempurl%
\url{https://www.usenix.org/conference/osdi12/technical-sessions/presentation/yuan}
\showURL{%
\tempurl}


\bibitem[Yuan et~al\mbox{.}(2012b)]%
        {yuan2012improving}
\bibfield{author}{\bibinfo{person}{Ding Yuan}, \bibinfo{person}{Jing Zheng},
  \bibinfo{person}{Soyeon Park}, \bibinfo{person}{Yuanyuan Zhou}, {and}
  \bibinfo{person}{Stefan Savage}.} \bibinfo{year}{2012}\natexlab{b}.
\newblock \showarticletitle{Improving software diagnosability via log
  enhancement}.
\newblock \bibinfo{journal}{\emph{ACM Transactions on Computer Systems (TOCS)}}
  \bibinfo{volume}{30}, \bibinfo{number}{1} (\bibinfo{year}{2012}),
  \bibinfo{pages}{1--28}.
\newblock


\bibitem[Zhang et~al\mbox{.}(2021)]%
        {zhang2021static}
\bibfield{author}{\bibinfo{person}{Jialu Zhang}, \bibinfo{person}{Ruzica
  Piskac}, \bibinfo{person}{Ennan Zhai}, {and} \bibinfo{person}{Tianyin Xu}.}
  \bibinfo{year}{2021}\natexlab{}.
\newblock \showarticletitle{Static detection of silent misconfigurations with
  deep interaction analysis}.
\newblock \bibinfo{journal}{\emph{Proceedings of the ACM on Programming
  Languages}} \bibinfo{volume}{5}, \bibinfo{number}{OOPSLA}
  (\bibinfo{year}{2021}), \bibinfo{pages}{1--30}.
\newblock


\bibitem[Zhang and Ernst(2013)]%
        {zhang2013automated}
\bibfield{author}{\bibinfo{person}{Sai Zhang} {and} \bibinfo{person}{Michael~D
  Ernst}.} \bibinfo{year}{2013}\natexlab{}.
\newblock \showarticletitle{Automated diagnosis of software configuration
  errors}. In \bibinfo{booktitle}{\emph{2013 35th International Conference on
  Software Engineering (ICSE)}}. IEEE, \bibinfo{pages}{312--321}.
\newblock


\bibitem[Zhang and Ernst(2014)]%
        {zhang2014configuration}
\bibfield{author}{\bibinfo{person}{Sai Zhang} {and} \bibinfo{person}{Michael~D
  Ernst}.} \bibinfo{year}{2014}\natexlab{}.
\newblock \showarticletitle{Which configuration option should I change?}. In
  \bibinfo{booktitle}{\emph{Proceedings of the 36th international conference on
  software engineering}}. \bibinfo{pages}{152--163}.
\newblock


\bibitem[Zhang and Ernst(2015)]%
        {zhang2015proactive}
\bibfield{author}{\bibinfo{person}{Sai Zhang} {and} \bibinfo{person}{Michael~D
  Ernst}.} \bibinfo{year}{2015}\natexlab{}.
\newblock \showarticletitle{Proactive detection of inadequate diagnostic
  messages for software configuration errors}. In
  \bibinfo{booktitle}{\emph{Proceedings of the 2015 International Symposium on
  Software Testing and Analysis}}. \bibinfo{pages}{12--23}.
\newblock


\bibitem[Zhao et~al\mbox{.}(2017)]%
        {zhao2017log20}
\bibfield{author}{\bibinfo{person}{Xu Zhao}, \bibinfo{person}{Kirk Rodrigues},
  \bibinfo{person}{Yu Luo}, \bibinfo{person}{Michael Stumm},
  \bibinfo{person}{Ding Yuan}, {and} \bibinfo{person}{Yuanyuan Zhou}.}
  \bibinfo{year}{2017}\natexlab{}.
\newblock \showarticletitle{Log20: Fully automated optimal placement of log
  printing statements under specified overhead threshold}. In
  \bibinfo{booktitle}{\emph{Proceedings of the 26th Symposium on Operating
  Systems Principles}}. \bibinfo{pages}{565--581}.
\newblock


\bibitem[Zhong et~al\mbox{.}(2024)]%
        {zhong2024automated}
\bibfield{author}{\bibinfo{person}{Renyi Zhong}, \bibinfo{person}{Yichen Li},
  \bibinfo{person}{Jinxi Kuang}, \bibinfo{person}{Wenwei Gu},
  \bibinfo{person}{Yintong Huo}, {and} \bibinfo{person}{Michael~R Lyu}.}
  \bibinfo{year}{2024}\natexlab{}.
\newblock \showarticletitle{Automated Defects Detection and Fix in Logging
  Statement}.
\newblock \bibinfo{journal}{\emph{arXiv preprint arXiv:2408.03101}}
  (\bibinfo{year}{2024}).
\newblock


\bibitem[Zhou et~al\mbox{.}(2021)]%
        {zhou2021confinlog}
\bibfield{author}{\bibinfo{person}{Shulin Zhou}, \bibinfo{person}{Xiaodong
  Liu}, \bibinfo{person}{Shanshan Li}, \bibinfo{person}{Zhouyang Jia},
  \bibinfo{person}{Yuanliang Zhang}, \bibinfo{person}{Teng Wang},
  \bibinfo{person}{Wang Li}, {and} \bibinfo{person}{Xiangke Liao}.}
  \bibinfo{year}{2021}\natexlab{}.
\newblock \showarticletitle{Confinlog: Leveraging software logs to infer
  configuration constraints}. In \bibinfo{booktitle}{\emph{2021 IEEE/ACM 29th
  International Conference on Program Comprehension (ICPC)}}. IEEE,
  \bibinfo{pages}{94--105}.
\newblock


\bibitem[Zhu et~al\mbox{.}(2015)]%
        {zhu2015learning}
\bibfield{author}{\bibinfo{person}{Jieming Zhu}, \bibinfo{person}{Pinjia He},
  \bibinfo{person}{Qiang Fu}, \bibinfo{person}{Hongyu Zhang},
  \bibinfo{person}{Michael~R Lyu}, {and} \bibinfo{person}{Dongmei Zhang}.}
  \bibinfo{year}{2015}\natexlab{}.
\newblock \showarticletitle{Learning to log: Helping developers make informed
  logging decisions}. In \bibinfo{booktitle}{\emph{2015 IEEE/ACM 37th IEEE
  International Conference on Software Engineering}}, Vol.~\bibinfo{volume}{1}.
  IEEE, \bibinfo{pages}{415--425}.
\newblock


\end{thebibliography}










\end{document}